\documentclass[12pt]{article}
\usepackage[top=1in,bottom=1in,left=1in,right=1in]{geometry}
\usepackage{amsmath, amsthm, amssymb}
\usepackage{setspace}
\usepackage{booktabs}
\usepackage{colortbl}
\usepackage{xcolor}
\usepackage{natbib}
\usepackage[hyphens]{url}
\usepackage[hidelinks]{hyperref}
\usepackage{graphicx}
\usepackage{pdflscape}
\usepackage{dcolumn}
\usepackage{pdfpages}
\usepackage{abstract}
\usepackage{titlesec}
\usepackage{enumerate}
\usepackage{svg}

\graphicspath{{Figures/}}

\defcitealias{ACTUS}{ACTUS}
\defcitealias{DoddFrank2010}{Dodd-Frank Wall Street Reform and Consumer Protection Act (2010)}

\titleformat{\paragraph}{\normalfont\normalsize\bfseries}{\theparagraph}{1em}{}
\titlespacing*{\paragraph}{0pt}{3.25ex plus 1ex minus .2ex}{1.5ex plus .2ex}

\begin{document}

\title{The Feasibility of MBSs as \\ Decentralized Autonomous Organizations}

\author{Timothy Dombrowski \\
	Assistant Professor of Finance \\
	Department of Finance and Legal Studies \\
	University of Missouri--St. Louis \\
	\href{mailto:tdombrowski@umsl.edu}{tdombrowski@umsl.edu} \\
	\\
	and \\
	\\
	V. Carlos Slawson Jr.\\
	Professor of Finance \\
	Department of Finance \\
	Louisiana State University \\
	\href{mailto:cslawson@lsu.edu}{cslawson@lsu.edu} \\ \vspace{0em}
}

\date{
	%\today
	May 1, 2024\thanks{We are grateful for comments from participants at the 2024 UCF/JRER Current Issues in Real Estate Symposium, the 2024 American Real Estate Society Spring Conference, and the University of Tulsa's Center for Real Estate Studies Seminar. We especially appreciate the insightful comments from Cayman Seagraves, Sergio G\'{a}rate, Charles Hilterbrand, and Tetyana Balyuk.} \\
	\vspace{2em} This draft is the SSRN pre-print as of May 1, 2024. \\
	First Draft: December 1, 2023. \\
	Final Version was published in \href{https://law-store.wolterskluwer.com/s/product/real-estate-finance/01tG000000LtrU8IAJ}{\emph{Real Estate Finance}} Summer 2024. \\ 
	\vspace{1em}
	To cite the final published version, please use the following citation: \\
	\vspace{0.5em}
	\parbox{\textwidth}{
		\hangindent=1.5em
		Dombrowski, T.P. and Slawson Jr., V.C. (2024), ``The Feasibility of Mortgage-Backed Securities as Decentralized Autonomous Organizations.'' \emph{Real Estate Finance}, 41:1, 1–20.
		}
	}

\maketitle

\newpage

\begin{center}
	\Large\textbf{The Feasibility of MBSs as \\ Decentralized Autonomous Organizations}
\end{center}\
\

\begin{abstract}
	\vspace{-1.5em}
	\indent
	
	Can the general structure of a mortgage-backed security (MBS) contract be programmatically represented through the use of decentralized autonomous organizations (DAOs)? Such an approach could allow for the portfolio of loans to be managed by investors in a trustless and transparent way. The focus and scope of this paper is to explore the potential for applying the tools of modern fintech, such as asset tokenization, smart contracts, and DAOs, to reconstruct traditional structured products that have a greater degree of transparency and traceability. MBS investors face considerable value uncertainty as time increases between the actual occurrence (or non-occurrence) of cash flows and subsequent reporting. Given that an MBS is a financial contract, it should be expressible logically using the Algorithmic Contract Types Unified Standards (ACTUS). Since each underlying mortgage in an MBS derives its cash flows in a prescribed way over the life of the contract, implementation on a public blockchain could enable real-time ratings systems, improving market efficiency. We explore the potential for creating formal algorithmic designs of MBS-DAOs that incorporate individual mortgages, the underlying real estate assets (collateral), and any loan guarantees.
	
	\vspace{0.1in}
\end{abstract}

%\pagebreak
%
%\large
%\tableofcontents

\pagebreak

\large
\setstretch{1.5}

%%%%%%%%%%%%%%%%%%%%%%%%%%%%%%%%%%%%%%%%%%%%%%%%%%%%%%%%%%%%%%%
% INTRODUCTION
%%%%%%%%%%%%%%%%%%%%%%%%%%%%%%%%%%%%%%%%%%%%%%%%%%%%%%%%%%%%%%%

\section{Introduction}\label{intro}

Early mortgage-backed securities (MBSs) in the latter half of the 18th century resulted in borrowers defaulting months before overseas investors were made aware \citep{Jong2023}. During the subprime mortgage crisis in the early 21st century, the time lag between default and MBS investor awareness was revealed to have not been much better than it was 250 years ago.

The advent of blockchain technology promises the creation of more efficient and transparent financial systems.  This paper aims to explore the potential for using blockchain-based smart contracts to build MBSs that should provide a greater degree of transparency and market efficiency.  Is it possible to build a system where MBS investors would know immediately whether or not the cash flows on the underlying mortgages had occurred? What are the implications around borrower and investor privacy if such contractual information were to be tracked on a public blockchain?

An MBS bundles together a portfolio of mortgages and issues securities that derive their value from the cash flows of the underlying loans. A hypothetical MBS Decentralized Autonomous Organization (MBS-DAO) would replicate the structure of a traditional MBS, but with a minimal lag between the occurrence of cash flows (or their non-occurrence) and investor awareness. MBS-DAO investors would immediately have access to information about the performance of the underlying mortgages.

Following the invention of Bitcoin \citep{Nakamoto2008} and the popularization of blockchain technology in the decade after that, a new type of business formation has emerged and has started to gain legal recognition. These have become known DAOs (pronounced `dows'). DAOs leverage blockchain-based smart contracts to establish the rules of governance for the entity. Despite the often misleading inclusion of the `decentralized' term, it continues to be used in reference to these blockchain-based entities. 

This paper aims to explore the potential for using blockchain-based smart contracts to build structured financial products that can provide a greater degree of transparency and market efficiency. How close are we to being able to create MBS-DAOs? What are the major concerns, issues, problems, and drawbacks? What are the major hesitations of investors? 

In the blockchain space, the legalities are often unclear. Yet, in the U.S., at least two states, Wyoming and Oklahoma, have seen projects experiment with the tokenization of real estate. Several states have even passed laws that allow DAOs to register as LLCs. This paper includes a discussion of some real estate focused DAO-LLCs that are registered in Wyoming (Section~\ref{tokenization}). Appendix~\ref{wyodaoappend} includes the first page of Wyoming's DAO-LLC registration form, which is very similar to the paperwork of other business structures (e.g., LLCs, partnerships, and corporations).

Can the general structure of an MBS contract be programmatically represented through the use of DAOs? The remainder of the paper is structured as follows:  Section~\ref{implementation} discusses the contractual components of an MBS-DAO. Section~\ref{actusstandards} reviews the relevant ACTUS standards for algorithmic financial contracts. To conclude, Section~\ref{discussion} includes a discussion of several considerations for implementing such a system. 

%%%%%%%%%%%%%%%%%%%%%%%%%%%%%%%%%%%%%%%%%%%%%%%%%%%%%%%%%%%%%%%%%%
% BLOCKCHAIN IMPLEMENTATION
%%%%%%%%%%%%%%%%%%%%%%%%%%%%%%%%%%%%%%%%%%%%%%%%%%%%%%%%%%%%%%%%%%

\section{Blockchain Implementation}\label{implementation}

In this section, the structure of an MBS investment is decomposed into three key components: (1) the underlying real estate properties, or loan collateral, (2) mortgage contracts secured by those properties, and (3) securitization of cash flow rights from mortgage pools. These three pieces are related to the blockchain concepts of real estate tokenization (Section~\ref{tokenization}), mortgage smart contracts (Section~\ref{mortgage}), and DAO-managed mortgage portfolios (Section~\ref{tranching}). Along each of these steps, the discussion will include examples of existing projects and considerations for applying these tools on various blockchains. 

\subsection{Real Estate Title Tokenization}\label{tokenization}

The first step towards designing an MBS-DAO is the process of tokenizing the underlying real estate assets. This is effectively the process of creating a digital asset to represent the ownership of the property. At the core of this idea is the concept of a non-fungible token (NFT), which simply means that each individual token has some degree of uniqueness to it. In some cases, this uniqueness can be as small as a serial number;\footnote{See NBA Top Shot (\url{https://nbatopshot.com/}) and NFL All Day (\url{https://nflallday.com/)}.} however, in other cases, individual NFTs can be much more distinctly unique.\footnote{For example, see Beeple's NFT titled `Everydays: The First 5000 Days', which sold for \$69.3 million in 2021 at Christie's \citep{Crow2021}.} 

With real estate assets, every property is a unique monopoly on a particular location. Thus, an NFT could represent some ownership of an individual property or unit. However, fungible tokens can also be relevant in this real estate tokenization context. For example, if a property owner wished to fractionalize ownership in a specific real estate property or investment, one could issue fungible tokens (shares) for that project.\footnote{See \url{https://www.pacaso.com/}, \url{https://realt.co/} and \url{https://fraction.estate/} for some examples of projects aimed at enabling fractional ownership of real estate assets.} If the property's title is tokenized, this would be done via a smart contract that acquires the NFT, and then issues shares representing the fractional claims. This introduces the topic of joint ownership and collective governance over the property, which is another potential application for DAOs that will be explored in this section.

Some of the earliest ideas on this topic focused on tokenizing virtual real estate.\footnote{For example, the Decentraland project raised over \$26 million (over 86,000 ETH) in a little over an hour during its 2017 initial coin offering for its MANA token \citep{Higgins2017}. The fungible MANA tokens act as the currency of the Decentraland virtual world that was launched in 2020. In 2018, MANA-holders were able to purchase non-fungible LAND tokens, which represent ownership of the virtual real estate parcels.} However, the tokenization of phsyical real estate requires a legal enforcement component recognized by the jurisdiction in which the property resides. This is demonstrated by the disclaimers within the NFT metadata of the two tokenization projects that are explored in this section. For a wider survey of real estate tokenization projects in the U.S. and around the world, see \citet{Saari2022} and \citet{Dombrowski2024a}. 

The remainder of this section will explore two U.S.-based projects experimenting with the development of processes for real estate tokenization. Section~\ref{citydao} focuses on CityDAO, which launched on the Ethereum blockchain in 2021, and Section~\ref{gokey} reviews GoKey DAO, which is a Cardano-based project that also launched in 2021. Beyond these two examples of DAO-managed real estate projects, \citet{Nagl2023} explore two other platforms focused on the tokenization of commercial real estate, Propy and Elysia.\footnote{See \url{https://propy.com/} and \url{https://www.elysia.land/}, for more detail on those projects.} For more detail on the concept of tokenized real estate, see \citet{Chainlink2023}. 

\subsubsection{CityDAO}\label{citydao}

One U.S.-based project exploring with the tokenizion of real estate assets on a blockchain is CityDAO.\footnote{\url{https://citydao.io}} In July 2021, CityDAO registered as a ``limited liability autonomous organization'' under the Wyoming DAO Law\footnote{See Section~\ref{wyolaw} for more on this legislation.} and then issued Citizenship NFTs, which allowed investors to finance the DAO with seed funding in ETH.\footnote{See \url{https://etherscan.io/token/0x7eef591a6cc0403b9652e98e88476fe1bf31ddeb} to view details about the CityDAO Citizenship NFTs, read its contract, and track the tokens on the Ethereum blockchain. Or see \url{https://opensea.io/collection/cdao} for the project's OpenSea page where one can view and transact the tokens.} These NFTs serve as governance tokens for CityDAO and allow holders to propose DAO actions and vote on proposals.\footnote{See \url{https://snapshot.org/\#/daocity.eth} for details about CityDAO proposals and voting records.}

Also during late 2021, CityDAO acquired a 40-acre plot of vacant land in northern Wyoming (Parcel Zero). Then in May 2022, CityDAO established Parcel Zero, LLC,\footnote{Parcel Zero, LLC, is not registered as a DAO-LLC, just as a traditional LLC in Wyoming.} which is listed as the property owner as of late 2023. CityDAO divided the parcel into plots and issued NFTs that represent each plot.\footnote{See \url{https://etherscan.io/token/0x90384e76b6b3ddb47396ff85144819ded148900d} to view details about the Parcel-0 tokens, read its contract, and track the tokens on the Ethereum blockchain. Or see \url{https://opensea.io/collection/citydao-parcel-0} for the project's OpenSea page.} Although the main webpage for the Parcel Zero project suggests that there are 10,000 plots,\footnote{\url{https://www.citydao.io/parcels/parcel-0}} the Parcel Viewer tool shows only 6,369 plots, with various pathways marked as common areas.\footnote{\url{https://parcel-0.citydao.io/}} From blockchain-based data sources, Etherscan and OpenSea, their data shows only 5,453 NFTs issued.

The metadata of these Parcel Zero NFTs contains several descriptive characteristics about the particular plot that the NFT represents. These include several categorical descriptors, such as the region (five districts/subdivisions), elevation (upper, mid, or lower), and slope (flat-ish, shallow, steep, very steep, or extremely steep). Additionally, there are three numeric descriptors that each measure a weight for the terrain composition (e.g., 34\% gravel, 3\% rock, and 63\% vegetation). Another component of the token metadata is a general description of the token. Of note within that description is the following:
\begin{quote}
	``This NFT does not represent any land ownership.  This Parcel 0 NFT represents a license agreement between the Holder and Parcel Zero, LLC (a wholly-owned subsidiary of CityDAO, LLC), which grants the Holder certain use and governance rights over the property.'' -- CityDAO Parcel Zero NFT Metadata
\end{quote}

Also included within the metadata of these tokens are two IPFS links.\footnote{IPFS (or InterPlanetary File System) is a file sharing protocol that has become common for storing files and metadata associated with NFTs. See \url{https://ipfs.tech} for more info.} The first links to a KML file\footnote{KML is short for Keyhole Markup Language. See \url{https://www.ogc.org/standard/kml/} for more info.} that contains a copy of the token metadata along with geographic data that can be loaded into Google Earth. Most Parcel Zero plots are squares, but some are irregularly shaped plots around the edge of the parcel and adjacent to the pathways. The second link directs to the Parcel 0 License Agreement, which specifies the use and governance rights over the property. The use of IPFS allows for most of the data to be stored \emph{off-chain} with just the IPFS links (hashes) being included \emph{on-chain}. This minimizes the required block space to mint and store the NFTs. 

As of late 2023, the CityDAO Citizenship NFTs have experienced over 4,400 ETH\footnote{Since these NFTs operate on the Ethereum blockchain, it is typical to measure price and volume in ETH. However, since stablecoins also exist on the Ethereum blockchain, it is technically possible to transact these tokens with a digital version of USD. Section~\ref{stable} discusses this topic of stablecoins in more detail.} of trading volume, as measured by OpenSea. Of that trading volume, the vast majority occurred during the first six months after minting (Nov. 2021 -- May 2022) with much less during the 18 months following that. Of note within the contract data for both sets of CityDAO NFTs is a royalty that charges a 10\% fee on secondary market transactions, which gets directed to the CityDAO Treasury Contract.\footnote{\url{https://etherscan.io/address/0x60e7343205C9C88788a22C40030d35f9370d302D}} This treasury portfolio holds digital assets (primarily WETH and USDC)\footnote{See Section~\ref{stable} for more on stablecoins, such as USDC, and wrapped assets, such as WETH.} as well as the Parcel Zero property.

The CityDAO Parcel-0 NFTs have experienced only 99 ETH of trading volume since their issuance in May 2022. However, unlike the Citizenship NFTs, which peaked in volume and price within the first few months of trading, the Parcel-0 NFTs saw a major spike in both price and volume at the end of 2022. This can be explained away by examining the public blockchain data. The outlier transactions driving those spikes are verifiable wash sales that account for nearly 48 ETH in volume across 4 transactions.\footnote{See Appendix~\ref{citydaoappend} for documentation of these transactions and an example of psuedonymous account identities on Ethereum.} Thus, the actual Parcel-0 trading volume since issuance is closer to 50 ETH.

A key difference between the CityDAO Citizenship NFTs and the Parcel Zero NFTs is the token standard used to generate the tokens. Although both are issued on the Ethereum blockchain, the Citizenship NFTs use the ERC-1155 standard, and the Parcel Zero NFTs use the ERC-721 standard.\footnote{The most common token standards on Ethereum are ERC-20 (fungible tokens), ERC-721 (NFTs), and ERC-1155 (both). See \url{https://ethereum.org/en/developers/docs/standards/tokens/} for more details.} The former standard enabled the creation of three different tiers to the Citizenship NFTs with varying quantities: 10,000 CityDAO Citizens, 50 Founding Citizens, and 1 First Citizen. The 1-of-1 First Citizen NFT was auctioned for 6.52 ETH ($\approx$ \$20K) in August 2021,\footnote{\url{https://city.mirror.xyz/7rtqWC6DBNHbIt87xCPUqkmFcw_3u_ZAK6vahn2IhOI}} and later sold for 39 ETH ($\approx$ \$111K) in May 2022.\footnote{\url{https://etherscan.io/tx/0x9ba46a56b1e169218200130765e1cbbe29e63d5fcfe6f47d6aa6c8acf3c84ecb}} In other words, the 10,000 Citizenship NFTs are effectively fungible within that tier, but the First Citizen NFT is a unique digital asset. See the CityDAO Blog for more details on the differences between the tiers.\footnote{\url{https://city.mirror.xyz/}}

In regard to legally-recognized claims of ownership, holders of the CityDAO Citizenship NFTs could potentially argue for some pass-through ownership of some share of the Parcel-0 property. This is because the Citizenship NFTs represent governance tokens in the CityDAO LLC, which is to sole owner of the Parcel Zero LLC. Since the latter entity is only registered as a traditional LLC, it is unclear whether ownership of the Parcel-0 NFTs would carry any legal claim of ownership. Given the clear verbiage of the NFT descriptions that they do not represent any claim of land ownership, the usage and governance rights conveyed by the NFTs relate more to those of easements or licenses, rather than any actual legal claim of title.

As of April 2024, the CityDAO project appears to be on the verge of dissolution. The project founder, Scott Fitsimones, announced that he would be stepping down in July 2024 and transitioning control of the DAO's assets to new leadership.\footnote{\url{https://forum.citydao.io/t/citydao-transition/2376}} In particular, the transition plan suggests that the Wyoming LLC component of the DAO may be shut down and that its real estate assets would either be transfered to new leadership or sold. It is currently unclear how a dissolution would be resolved.

\subsubsection{GoKey DAO}\label{gokey}

Another U.S.-based project focused on tokenizing real estate assets is GoKey DAO.\footnote{\url{https://gokey.finance/}} This project is built on the Cardano blockchain, and similarly seeks to experiment with DAO-based governance of real estate assets. As with CityDAO, GoKey is registered as a DAO-LLC in the state of Wyoming. The governance tokens of the GoKey DAO, GOKEY, were issued in part through an Initial Stake Pool Offering (ISPO), which is a novel mechanism for fundraising that is enabled by Cardano's proof-of-stake consensus algorithm.\footnote{See \citet{Vasile2021} for more info about ISPOs and other crypto funding models, such as Initial Coin Offerings (ICOs).} The blockchain records on Cardanoscan show that the tokens were created in June 2021 and have a maximum supply of 4.5 billion GOKEY.\footnote{The token's authenticity can be verified by its `Token Policy Id', which is the last string in \url{https://cardanoscan.io/tokenPolicy/c7dcfa416c127f630b263c7e0fe0564430cfa9c56bba43e1a37c6915}.} This was roughly one year prior to the ISPO, which began in June 2022.\footnote{For more info on GoKey DAO's ISPO, see \url{https://medium.com/@gokeyfinance/announcing-the-gokey-ispo-and-staking-rewards-5d920cc1eed2}.}

Without going into all the detail about the GoKey ISPO, this process effectively substitutes the standard Cardano staking rewards with GOKEY tokens. More specifically, holders of Cardano's ADA can delegate their stake to the GoKey Stake Pool,\footnote{\url{https://pooltool.io/pool/c3c27acc9633f75202050d99f475178cff31cfde6d059696cc2d97f9}} which helps validate transactions on the Cardano blockchain. Unlike other stake pools that typically charge between 0--5\% in fees, an ISPO stake pool is set up to retain 99+\% of the staking rewards that the pool generates. This inflow of ADA to the GoKey DAO provides a sustained source of liquidity for the DAO. Then in return for foregoing their ADA staking rewards, delegators to the pool receive GOKEY tokens instead. 

\begin{figure}[htbp]
	\centering
	\includegraphics[width=\linewidth]{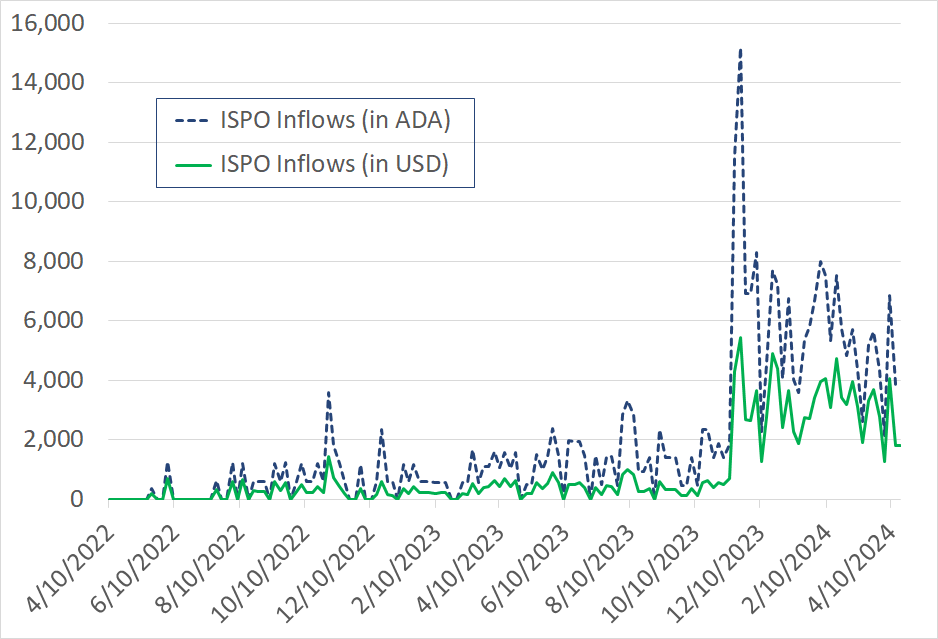}
	\caption{GoKey ISPO Inflows by 5-Day Epoch in ADA and USD. The staking data is from Cardano Pool Tool (\url{https://pooltool.io/}), and the ADA price data is from Yahoo Finance (\url{https://finance.yahoo.com/}). As of April 2024, the total ISPO inflows are approximately 290,000 ADA or \$135,000.}
	\label{gokeyplot}
\end{figure}

Due to the non-custodial nature of staking on the Cardano blockchain, this effectively allows investors to receive GOKEY tokens without having to transfer any funds to the project's developers. Since the beginning of the ISPO in mid-2022, the GoKey Stake Pool has generated more than 290,000 ADA, which amounts to roughly \$135,000 if each ADA payment was converted to USD immediately upon receipt. Figure~\ref{gokeyplot} depicts the \emph{epoch tax} (or ISPO inflows) in both ADA and USD for every 5-day epoch since the beginning of the GoKey Stake Pool. Of note is epoch 447 (ended November 11, 2023), in which the GoKey Stake Pool had a substantial increase in stake. 

In March 2022, GoKey minted a Warranty Deed NFT on the Cardano blockchain that represents a proof-of-concept for the tokenization of real estate records.\footnote{\url{https://pool.pm/asset18q93muz5ne5l62ac7xmvuzwzyfn3z6ck43gscc}} This NFT includes an IPFS link to an image file of a document that contains details about the property, which is located in Tulsa, Oklahoma. Also included within that image file is a QR code that links to a General Warranty Deed from 2019 for the property.

Then in March 2023, GoKey minted a Property Data Record NFT that contains additional information and links relating to that Tulsa property.\footnote{\url{https://pool.pm/asset1r44njx9wxsewlhs674xaedxvhy46cuz5n0vld3}} Similar to the CityDAO Parcel Zero NFTs, the metadata and associated files are hashed into IPFS storage before being timestamped into the blockchain. Within the metadata are several fields, including the property's street address, legal description, assessor's property number, country, state, county, land area (acreage), zoning code, links to the zoning map, tax assessor, and county clerk websites, as well as a legal disclaimer, which states the following:
\begin{quote}
	``This NFT is a digital asset which verifiably and immutably stores and records legally recognized
	physical and legal data and facts regarding this property on the Cardano blockchain. GoKey has verified this user provided property metadata and it is believed to be accurate as of the NFT mint timestamp. GoKey does not warrant or guarantee its accuracy. Interested parties
	must perform due diligence and must not rely solely on this data.''
\end{quote}

If we examine the on-chain records for the Warranty Deed NFT,\footnote{\url{https://cardanoscan.io/token/asset18q93muz5ne5l62ac7xmvuzwzyfn3z6ck43gscc}} the Cardanoscan shows eight total transactions as of the end of 2023. The first of these is the mint transaction in March 2022. Of the remaining seven transactions, only two actually involve transferring the NFT to another wallet.\footnote{The other five transactions include the NFT within the transaction; however, these are intra-wallet transfers. One of which is the delegation of that wallet to the GoKey Stake Pool in June 2022.} In early March 2023, before the minting of the Property Data Record NFT, the Warranty Deed NFT was tranferred to a new wallet, which was delegated to a different Cardano stake pool. It was transferred again in late November 2023 back to the wallet that originally minted the NFT.

When researching the Property Data Record NFT, the blockchain data shows two transactions as of late 2023.\footnote{\url{https://cardanoscan.io/token/asset1r44njx9wxsewlhs674xaedxvhy46cuz5n0vld3}} The first was the mint of the token in March 2023 by the same wallet that minted the Warranty Deed NFT, and the second was just an intra-wallet transfer in June 2023.

If we compare those on-chain records with the county clerk records, which are linked to in the Property Data Record NFT's metadata, the 2019 General Warranty Deed matches the document linked to in the Warranty Deed NFT. More recently, there is an October 2023 General Warranty Deed in the county records that does not coincide with any associated transfer of either NFT.

In summary, these two projects are some early ventures into the tokenization of real estate assets on public blockchains. In both cases, the token metadata includes a disclaimer noting that the tokens do not convey any legal title to the underlying real estate. Given that any claim of representing legal title would require some degree of legal recognition from a local government, this step remains a major hurdle for real estate tokenization projects.

\subsection{Mortgage Smart Contract}\label{mortgage}

At this point, let us assume that there exists some form of legally recognized tokenization process for real estate assets. If so, the development of a smart-contract-based mortgage is similar to that of any other collateralized loan. Some examples of Ethereum-based applications for collateralized lending are Aave, Compound, and MakerDAO.\footnote{See \url{https://aave.com/}, \url{https://compound.finance/}, and \url{https://makerdao.com/}, respectively.} In general, these protocols allow for users to deposit collateral in the form of digital assets, and then borrow against that collateral. Given the volatile nature of most digital assets, these protocols often require substantial overcollateralization. 

With real estate assets as collateral, or at least digital representations of such, the legal recognition of such tokenization process is a crucial step toward applying such DeFi lending protocols to create a smart-contract-based mortgage. In other words, if the value of the collateral can suddenly drop to zero due to the local jurisdiction revoking the legitimacy of the property's NFT, then this could pose a substantial risk for potential lenders. Perhaps an analogy for this would be a traditional case where a homebuyer purchases a home with an associated mortgage, and later discovers that one of the deeds in the chain of title is actually fraudulent or that there is some other cloud to the title. The mitigation of this risk may be addressable by title insurance companies.

For example, in the GoKey example from Section~\ref{gokey}, the metadata of the Property Data Record NFT contains several items that are part of a title search process. For example, Item 11 is a link to the county clerk's website for the property, which includes links to the deeds back to 1999. This research into the historical deeds and chain of title is a major component to the title search process. The tokenization of such records, as in the GoKey example, can be seen as reducing the cost of such verification for any future transfers of the property. However, the determination of whether or not transfers of such tokens can be legally recognized as formal transfers of ownership remains unsettled. See \citet{Spielman2016}, \citet{Ewendt2018}, \citet{Koronczok2019}, and \citet{Freyermuth2023} for more detailed legal discussions on applying blockchain to title insurance processes.

Beyond the collateral component of a mortgage, there are two other important contractual elements to discuss: the annuity/cash flow structure and any loan guarantees. In the U.S., the 30-year fixed rate mortgage has become standard for most residential mortgages. With a fixed rate, the cash flow structure is deterministically set at loan origination. This can be encoded into a smart contract fairly easily. However, more complex loan types, such as adjustable rate mortgages (ARMs) or hybrid ARMs, can incorporate market rates into the loan terms creating dynamic payment amounts. To implement these with smart contracts would likely require the use of oracles.\footnote{See Section~\ref{oracles} for a discussion of smart contract oracles and the ability to make smart contracts contingent on off-chain data/conditions.}

Lastly, the guarantee component of a mortgage relates to private mortgage insurance, as well as any government programs that may provide loan guarantees. This component will be discussed in more detail in Section~\ref{ceg}, and more generally, these three contractual elements (collateral, guarantees, and annuity) will be a major focus in Section~\ref{actusstandards}.

\subsection{Loan Pools, Tranching, and MBS-DAO Cash Flows}\label{tranching}

In order to begin pooling mortgages together to form a portfolio that generates cash flows for an MBS, there first must be an ability to transfer the cash flow rights of an individual loan. This can be achieved within a mortgage smart contract by tokenizing the roles (lender and borrower). In other words, rather than specifying the particular lender to receive the repayments from the borrower, the contract would issue a transferable NFT that designates where the cash flows should be directed. This would first be sent to the original lender, who could then sell it on the secondary market.

Alternatively, rather than issuing an NFT that receives the repayments, a more crowdsourced approach could be to issue fungible shares of the debt. However, this fractionalization of the debt potentially introduces governance issues in the event of non-performance of the loan. In other words, would there be enough coordination among the creditors to undergo the legal proceedings to regain ownership of the property in the real world? Depending on the strength of legal recognition of the property NFTs, much of this process is potentially automatable through the smart contract's pre-determined rules. However, there would likely need to be some coordination among the holders of the fractionalized debt shares.

Now, assuming that the cash flow rights to individual mortgages can be traded on a secondary market, an MBS-DAO (securitizer) would form portfolios of these tokens and issue tokens that receive cash flows derived from the underlying loans. This could be achieved in a centralized way where a traditional business entity acquires and custodies the NFTs that receive the mortgage debt cash flows. Alternatively, this could be achieved through the creation of a DAO that is managed algorithmically by its investors. This MBS-DAO would be a smart contract that issues governance tokens to investors that contribute seed funds and then acquires a portfolio of mortgage debt NFTs.

This hypothesized MBS-DAO would encounter a multitude of key governance decisions at its origination. At the core of any DAO is the responsibility of managing a portfolio of digital assets. Thus, the contract developers would need to encode rules around contributions and redemptions, as well as setting initial parameters for the investment strategy. Some examples for those parameters are the criteria for what debt NFTs to acquire or what other assets the DAO might hold, such as stablecoins or stablecoin-related products.\footnote{See Section~\ref{stable} for more on stablecoins.} After the DAO is funded, those parameters could be adjusted through the encoded governance system.

For the most part, any major governance decision could ultimately be made by rolling the DAO into a new contract. However, this would likely require a very high degree of consensus to achieve. Perhaps a strategy for mitigating the risks of low turnout voting would be to have multiple classes of MBS-DAO tokens with varying degrees of voting rights in the DAO governance. The specifics of this governance process could be as complex as it is in modern corporate governance structures with potentially more layers of complexity that are enabled through the blockchain-based data structures.\footnote{See \citet{Yermack2017, Zhao2022, Arshadi2023, Goldberg2023} for more discussions and analysis on the impact of blockchains and DAOs on corporate governance.}

\subsubsection{LiquidFi}

Another company working on tokenizing real estate assets/records is LiquidFi.\footnote{\url{https://www.liquidfi.io/}} This Miami, Florida-based company is focused on tokenizing post-origination mortgages and MBSs, which requires collaboration with originators, investors, and servicers. In 2021, they were issued a patent for their system of managing loans and securities using a distributed ledger \citep{Ferreira2021}. According to \citet{Gaffney2023}, ``Liquid Mortgage [LiquidFi] has reduced MBS reporting from 55 days to 30 minutes on the Stellar blockchain.'' In addition to the sub-30-minute settlement time, their website suggests managment of over 6,000 loans with more than \$6 billion in unpaid principal balances and nearly \$1 billion in payments recorded, as of April 2024. On the compliance side, their site also notes the completion of a SOC 2 Type II Compliance audit.\footnote{For more on System and Organization Control (SOC) audits from the American Institute of Certified Public Accountants (AICPA), see \url{https://www.aicpa-cima.com/topic/audit-assurance/audit-and-assurance-greater-than-soc-2}.}

\subsection{MBS-DAO Structure}

\begin{landscape}
	\begin{figure}[htbp]
		\centering	
		\includegraphics[width=1.2\textwidth]{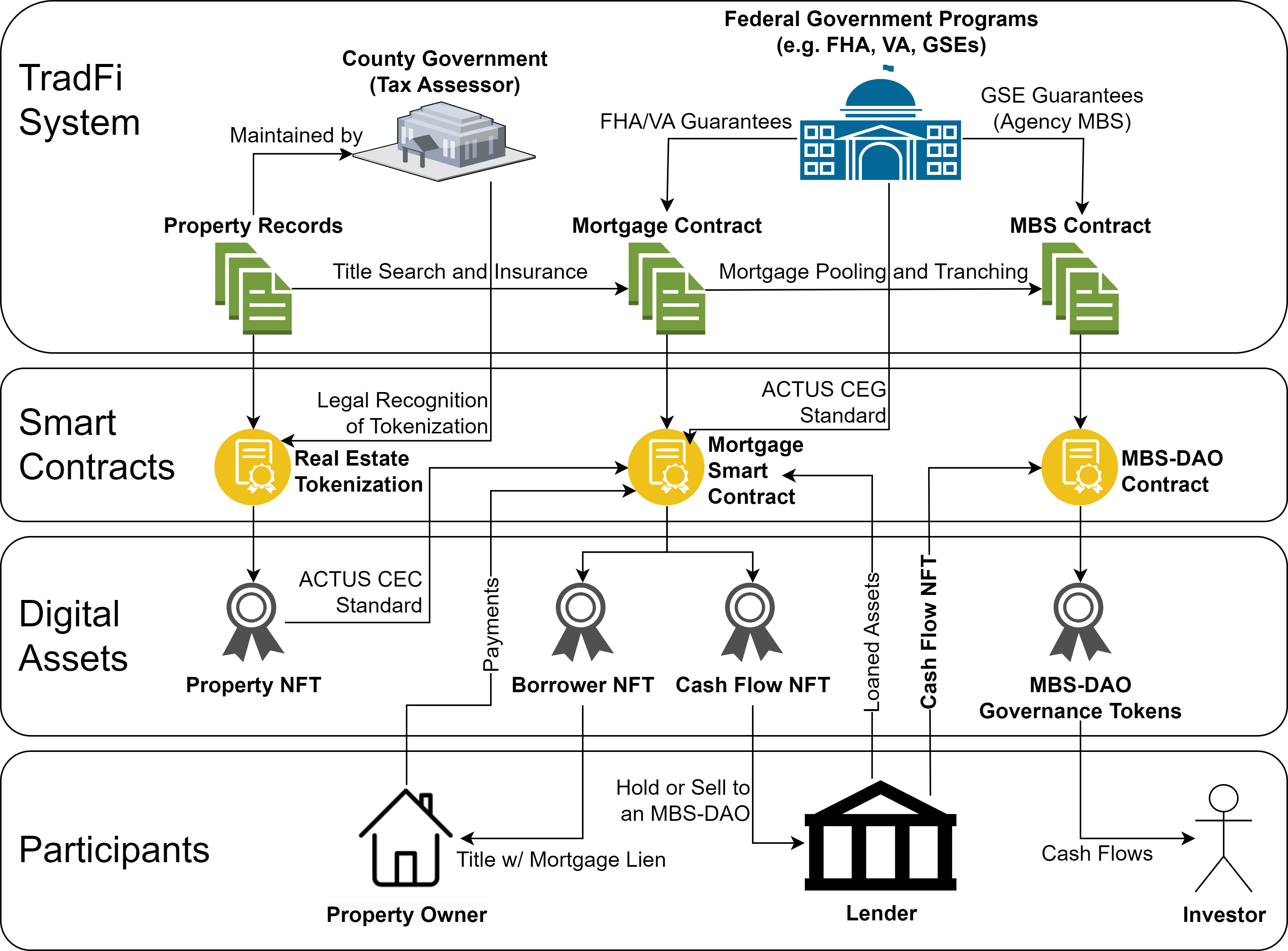}
		\caption{MBS-DAO Diagram -- The top portion depicts the general structure of the traditional MBS system, and the sections below visualize the proposed MBS-DAO structure.}
		\label{diagram}
	\end{figure}
\end{landscape}

To provide a cohesive design for this proposed MBS-DAO structure, Figure~\ref{diagram} on the next page depicts a mapping of the traditional MBS system (TradFi) to a blockchain-based system consisting of smart contracts and digital assets. This proposed MBS-DAO structure consists of three key smart contracts: (1)~tokenization contract, (2) mortgage contract, and (3) MBS-DAO contract. The diagram also depicts the cash flows between parties and contracts, as well as ACTUS contract standards, which will be the explored in Section~\ref{actusstandards}.

%%%%%%%%%%%%%%%%%%%%%%%%%%%%%%%%%%%%%%%%%%%%%%%%%%%%%%%%%%%%%%%%%%
% CONTRACT STANDARDS
%%%%%%%%%%%%%%%%%%%%%%%%%%%%%%%%%%%%%%%%%%%%%%%%%%%%%%%%%%%%%%%%%%

\section{ACTUS Contract Standards}\label{actusstandards}

In this section, the MBS-DAO structure will be explored through the lens of the ACTUS financial contract standards. The Algorithmic Contract Types Unified Standards \citepalias{ACTUS} are a set of standards maintained by the ACTUS Foundation. These standards are developed with a goal of developing ``consistent representation of financial instruments,'' or more specifically:
\begin{quote}
	``The goal of ACTUS is to break down the diversity in financial instruments into a manageable number of cash flow patterns – so called Contract Types (CT).'' -- The ACTUS Foundation 
\end{quote}

The ACTUS Taxonomy starts with a root split that separates \emph{Financial Contracts} from \emph{Credit Enhancement}.\footnote{See Appendix~\ref{actustaxappend} for the full taxonomy diagram, or go to \url{https://www.actusfrf.org/taxonomy}, which has all of the contract type definitions.} Although most of the specifications (leaf nodes) fall under the financial contracts branch, there are some credit enhancement standards that are relevant to the context of MBSs. This section will discuss several relevant standards for the construction of an MBS-DAO.

In short, the relevant building blocks for the creation of an MBS-DAO are the CEC (Collateral), CEG (Guarantees), and ANN (Annuity) standards. Additionally, the taxonomy includes two standards related to this MBS-DAO concept: the SCRMR (Securitization Market Risk) and SCRCR (Securitization Credit Risk). This distinction between market risk and credit risk relates to the tranching structure of MBS bonds and whether all underlying contracts are treated equal (market risk) or if losses first hit the most junior/residual tranches before waterfalling down to more senior tranches (credit risk).

The \emph{Credit Enhancement} standards, such as CEC and CEG, can be incoporated into the various \emph{Financial Contract} standards, such as ANN, SCRMR, and SCRCR. For example, if we consider an individual mortgage, then the cash flows are defined by the ANN component of the contract, and the collateral relationship of the underlying real estate securing the loan is defined by the CEC component. Any loan guarantees or insurance can be incorporated through the CEG component. These relations are depicted in Figure~\ref{diagram} with labeled arrows flowing into the mortgage smart contract.

Each of the first three standards (CEC, CEG, and ANN) has a rigorous technical specification available from ACTUS;\footnote{See \url{https://www.actusfrf.org/techspecs} for the detailed technical specifications, and \url{https://www.actusfrf.org/dictionary} for a dictionary of variable names and descriptions.} however, the two securitization standards are still listed as `planned' in late 2023. Rather than repeating the full specifications for the first three standards, this section will discuss each of these standards and its relevance to the construction of an MBS-DAO.

In 2021, the Office of Financial Research\footnote{The OFR was established under the U.S. Department of the Treasury as part of the \citetalias{DoddFrank2010}. The statutory goals of the agency are ``(1) collecting and standardizing data collections; (2) performing applied research and essential long-term research; and (3) developing risk measurement and monitoring tools.''} added the ACTUS data dictionary and standards to their Financial Instrument Reference Database (FIRD).\footnote{\url{https://www.financialresearch.gov/data/financial-instrument-reference-database/}} In addition to the ACTUS standards, the FIRD also includes the ISO20022\footnote{\url{https://www.iso20022.org/}} and FIX\footnote{\url{https://www.fixtrading.org/}} standards \citep{OFR2022}. These other two sets of standards each consider five types of asset classes: equities, debt, options, warrants, and futures.

\subsection{CEC: Collateral}\label{cec}

The ACTUS Taxonomy defines \emph{collateral} contracts as `credit enhancement', or more specifically:
\begin{quote}
	``Collateral creates a relationship between a collateral an obligee and a debtor, covering the exposure from the debtor with the collateral.''
\end{quote}

Within the technical specification for this standard, there are four \emph{events} in its contract schedule and three \emph{state variables} that are initialized at contract start. Additionally, the specification includes \emph{state transition functions} and \emph{payoff functions} for each event. Without going into all of the technical detail about the specification, the general idea of this standard is that it establishes the collateral component of a secured loan.

In the context of a mortgage, the collateral for the loan is the underlying real estate asset. This is legally enforced with a mortgage lien, which is the legal right of the creditor on a mortgage to claim the underlying collateral in event of default by the borrower. The specific laws regarding mortgage defaults, foreclosures, and evictions can vary across jurisdictions,\footnote{For example, recourse states vs. non-recourse states when it comes to deficiency judgments. See \citet{Ghent2011} and \citet{Nam2021} for more details on this distinction and related research.} which suggests that there may be variation in regard to the strength of this collateral component to a mortgage contract. 

In other words, even if a smart contract could automate the default and foreclosure processes in regard to recordkeeping, the physical nature of real estate necessitates a human element in the legal recognition of such digital records and the execution of eviction processes. Since we are considering structured products that involve pooling mortgages secured by properties across multiple jurisdictions, the nature of this collateral relationship may vary across those legal systems. Thus, the geographical composition of an MBS's loan pool can be a relevant factor when measuring portfolio-level risk.

As discussed in Section~\ref{tokenization}, real estate tokenization refers to the process of creating an NFT that holds some degree of legal claim to a real estate property. Following the design in Figure~\ref{diagram}, a leveraged real estate purchase would place this NFT in the custody of the mortgage smart contract. Then, just as with the tokenized lender role (Cash Flow NFT), the tokenized borrower role (Borrower NFT) would represent the property's title with a mortgage lien. Upon full performance of the mortgage terms, the contract releases the unencumbered Property NFT to the homeowner's self-custody wallet.

In the event of non-performance, the rules around delinquency and default would be programmed into the smart contract. For example, various mechanisms for outreach and monitoring could be triggered at multiple points of delinquency (e.g., 30-days, 60-days, etc.). At a certain point, the contract may then execute a programmatic default judgment, which would trigger a governance event where the lender(s) could initiate the foreclosure process.

\subsection{CEG: Guarantees}\label{ceg}

The \emph{guarantees} contract type is another `credit enhancement' defined as:
\begin{quote}
	``Guarantee creates a relationship between a guarantor, an obligee and a debtor, moving the exposure from the debtor to the guarantor.'' -- ACTUS Taxonomy
\end{quote}

The technical specificaton for this standard contains eight events in the contract schedule and seven state variables, along with functions for state variable transitions and payoffs. In short, this standard represents the inclusion of any loan guarantees or insurance accompanying the mortgage.

This standard is relevant within the MBS context given the various mortgage insurance and government programs that provide guarantees on mortgage loans and MBS. Some examples at the individual loan level are private mortgage insurance and Federal Housing Administration (FHA) loans. The latter of which are private loans that are regulated and insured by the FHA. At the MBS level, securitizers group loans into tranches based on default risk and issue securities that receive the cash flows from the loans in the tranche. At the most senior levels, these tranches are effectively guaranteed through the `explicit guarantee' of GSE conservatorship.

With the GSEs, there are multiple levels at which guarantees are relevant. At the highest level, their placement into conservatorship in 2008 took what had been known as an `implicit guarantee' to an `explicit guarantee' that the U.S. Government would not allow the companies to fail. Although there were efforts under the Trump administration to move control of the GSEs back to private ownership, such as allowing the companies to retain a larger share of their profits \citep{Ackerman2019}, those efforts have been unsuccessful. The Biden administration has signalled no interest in relinquishing control over the GSEs \citep{Ackerman2021}.

\subsection{ANN: Annuity}\label{ann}

The \emph{annuity} standard is defined by the ACTUS Taxonomy as:
\begin{quote}
	``Principal payment fully at Initial Exchange Date (IED) and interest plus principal repaid periodically in constant amounts till Maturity Date (MD). If variable rate, total amount for interest and principal is recalculated to be fully matured at MD.''
\end{quote}

The technical specification for this standard includes many more events and state variables. In total, there are 16 events in the contract schedule and 13 state variables, along with the state transition and payoff functions.

Within the context of U.S. mortgage markets, where 30-year fixed rate mortgages dominate, this contract specification will cover the majority of existing residential mortgages. This standard is even flexible enough to include simple adjustable-rate mortgages (ARMs); however, the more typical hybrid ARMs would fit into the LAX (Exotic Linear Amortizer) or NAX (Exotic Negative Amortizer) contract types.

When it comes to an individual mortgage, the credit enhancements of \emph{collateral} and \emph{guarantees} are inputs to the \emph{annuity} contract. More generally, the credit enhancements are potential elements of an annuity contract.

\subsection{SCRMR and SCRCR: Securitization Standards}\label{scr}

At the \emph{Securitization} level of the ACTUS Taxonomy, there are two branches: \emph{Credit} and \emph{Market}. Interestingly, both the securitization market risk (SCRMR) and securitization credit risk (SCRCR) standards fall under the market branch. The credit branch contains three standards that pertain to swap contracts: credit default swaps (CDSWP), total return swaps (TRSWP), and credit linked notes (CLNTE). Although those three credit standards could certainly be applied to derivatives of MBS contracts, we'll keep the focus on the first two standards that are more directly related to the MBS cash flows.

Since the two market securitization standards (SCRMR and SCRCR) are still labelled as `planned' in late 2023, they are not yet present in the technical specification document. However, each has a brief description in the taxonomy that will be explored within the context of an MBS-DAO.

The SCRMR standard refers to securitization market risk. This is defined by the ACTUS Taxonomy as:
\begin{quote}
	``Securitization contracts where all underlying contracts are treated equal. The buyer of a tranche gets a part of the cash-flows.''
\end{quote}

\noindent Whereas SCRCR (securitization credit risk) is defined as:
\begin{quote}
	``Securitization contracts where contracts are ranked according to credit default. The lower ranked tranches are hit by the first defaults. Only when the lowest tranches are wiped out, the next higher tranch is hit.''
\end{quote}

In regard to the distinction between these two standards, the former (SCRMR) refers to a simple tranching structure where an MBS buyer simply is entitled to proportional claim of the cash flows generated by the portfolio. Some degree of complexity to the tranching structure can be incorporated into this standard, such as separating cash flows into interest-only (IO) and principle-only (PO) tranches. In these cases, the cash flows can vary greatly from month to month with great complexity.

For example, if we consider the IO vs. PO tranching structure, the risks of default and prepayment have varying impacts on the two tranches. In a period of decreasing interest rates, borrowers are more likely to refinance a loan to take advantage of the lower rates. For MBS investors, the buyers of the PO tranche are relatively unaffected; however, the IO tranche experiences decreased cash flows.

More complex tranching structures, such as those that issue bonds with varying levels of seniority, fit more into the SCRCR standard. Since those MBS are more similar to fixed income obligations, the cash flows behave more like a credit product than a volatile market product, hence the distinction in standard names. An MBS-DAO could implement this standard to automate the cash flows of a waterfall tranching structure.

Both of the SCRMR and SCRCR standards refer to contracts derived from a securitized product. Thus, these standards apply more broadly than just MBS and could be applied to other asset-backed securities (ABS), such as auto loan ABS, credit card ABS, or student loan ABS. Each of those could merit a discussion of the various components (collateral, guarantees, etc.). However, mortgages have a distinct form of collateral, which is real estate assets. This provides an opportunity to incorporate other aspects of real estate fintech innovation, such as real estate tokenization.

%%%%%%%%%%%%%%%%%%%%%%%%%%%%%%%%%%%%%%%%%%%%%%%%%%%%%%%%%%%%%%%%%%
% DISCUSSION
%%%%%%%%%%%%%%%%%%%%%%%%%%%%%%%%%%%%%%%%%%%%%%%%%%%%%%%%%%%%%%%%%%

\section{Discussion}\label{discussion}

This section will explore several legal and technical considerations relating to the creation of an MBS-DAO as described in this paper. 

\subsection{Legal Considerations}\label{legal}

The legal complexities of real estate tokenization have already been noted throughout this paper, specifically within Sections~\ref{tokenization} and \ref{cec}. This section will discuss some other legal considerations along the process of constructing an MBS-DAO, such as the recognition of DAOs as legal business entities (Section \ref{legaldaos}), KYC/AML compliance (Section \ref{kycaml}), and stablecoin adoption/regulation (Section \ref{stable}).

\subsubsection{DAOs as Business Entities}\label{legaldaos}

To our knowledge, there are four U.S. states that have passed laws relating to the establishment of DAOs (or similarly defined entities) as legal business entities. These are Vermont, Wyoming, Tennessee, and Utah. At the federal level, there is currently a Senate bill that was proposed in 2022 with an entire section focused on DAOs, among other related topics such as stablecoins.

\paragraph{Vermont}\label{vtlaw}

In 2018, the Vermont General Assembly passed S.269 (Act 205),\footnote{\url{https://legislature.vermont.gov/bill/status/2018/S.269}} which relates to blockchain business development. Although this law does not specifically use the verbiage of DAOs, it does include a similar term that is described as a ``blockchain-based limited liability company (BBLLC).'' The law then specifies the requirements for such a BBLLC registration, including an explicit mention of its intention to be a BBLLC in its articles of organization.

In addition to this election to be a BBLLC, the law also requires several items to be included in the organization's operating agreement. Among these requirements are (1) whether the underlying blockchain ledger is ``fully decentralized or partially decentralized,'' (2) whether the ledger will be ``fully or partially public or private,'' (3) ``adopt voting procedures, which may include smart contracts,'' (4) ``how a person becomes a member of the BBLLC,'', and (5) ``specify the rights and obligations of each group of participants.''

Another relevant component of the law is a section on ``Consensus Formation Algorithms and Governance Processes.'' This section establishes a reasonableness standard to the ``algorithmic means for accomplishing the consensus process for validating records, as well as requirements, processes, and procedures for conducting operations, or making organizational decisions on the blockchain.''

\paragraph{Wyoming}\label{wyolaw}

In the 2021 legislative session for the state of Wyoming, the state passed SF0038, or the Wyoming DAO Law,\footnote{\url{https://www.wyoleg.gov/Legislation/2021/SF0038}} which established a legal definition for a DAO that recognizes the business formation as a ``limited liability autonomous organization'' (or LAO). Since then, the state has seen several DAOs register and begin operating, such as CityDAO and GoKey DAO.

After equating the DAO and LAO concepts, the law defines a DAO as:
\begin{quote}
	``(a) A decentralized autonomous organization is a limited liability company whose articles of organization contain a statement that the company is a decentralized autonomous organization as described in subsection (c) of this section.'' -- Wyoming DAO Law
	
	``(c) A statement in substantially the following form shall appear conspicuously in the articles of organization or operating agreement, if applicable, in a decentralized autonomous organization:
	
	NOTICE OF RESTRICTIONS ON DUTIES AND TRANSFERS
	
	The rights of members in a decentralized autonomous organization may differ materially from the rights of members in other limited liability companies. The Wyoming Decentralized Autonomous Organization Supplement, underlying smart contracts, articles of organization and operating agreement, if applicable, of a decentralized autonomous organization may define, reduce or eliminate fiduciary duties and may restrict transfer of ownership interests, withdrawal or resignation from the decentralized autonomous organization, return of capital contributions and dissolution of the decentralized autonomous organization.'' -- Wyoming DAO Law
\end{quote}

See Figure~\ref{wyodaoform} in Appendix~\ref{wyodaoappend} for the first page of the organization form for a DAO-LLC to register in the state of Wyoming. Of note in the form is question 8(a), which asks about the governance structure of the DAO in regard to whether it is member managed or algorithmically managed.

More recently in 2024, Wyoming passed SF0050, or the Wyoming Decentralized Unincorporated Nonprofit Association (DUNA) Act. Within this law, there are sections that explicitly deal with the ownership and transfer of real estate. For example, Section 17-32-105 states:
\begin{quote}
	``A decentralized unincorporated nonprofit association in its name may acquire, hold, encumber or transfer an estate or interest in real or personal property.'' -- Wyoming DUNA Law
\end{quote}

To facilitate the property ownership and any transfers by a Wyoming DUNA, the law requires the registration of a \emph{statement of authority as to real property} with the county clerk. This statement of authority specifies a person authorized to transfer real property held by the DUNA. Additionally, the statement of authority is required to include the ``action, procedure or vote'' that authorizes the person to make the transfer.

\paragraph{Tennessee}\label{tnlaw}

In 2022, the Tennessee General Assembly passed HB 2645,\footnote{\url{https://wapp.capitol.tn.gov/apps/BillInfo/Default.aspx?BillNumber=HB2645&GA=112}} which similarly establishes legal clarity for DAOs. One distinction of this law compared to the Wyoming DAO Law is that it refers to the business entity as a ``decentralized organization'', rather than a DAO. However, the DAO abbreviation is included within the list of allowed verbiage for explicitly denoting its status. Thus, the term ``autonomous'' is absent from the law, except for its implied usage within the DAO acronym.

Another distinction in wording between the Tennessee law and the other two state laws is that it makes use of the phrase ``distributed ledger technology'' rather than ``blockchain'' when describing the underlying system. Its definition for this phrase includes the term ``blockchain'' and is as follows:
\begin{quote}
	```Distributed ledger technology' means a distributed ledger protocol
	and supporting infrastructure, including blockchain, that uses a distributed,
	decentralized, shared, and replicated ledger, whether it be public or private,
	permissioned or permissionless, and that may include the use of electronic
	currencies or electronic tokens as a medium of electronic exchange'' -- Tennessee DAO Law
\end{quote}

\paragraph{Utah}\label{utlaw}

The 2023 Utah DAO Law is another state law providing a legal framework for DAOs to register as a business entity.\footnote{\url{https://le.utah.gov/~2023/bills/static/HB0357.html}} Within its definition for a DAO, the organization must be ``created by one or more smart contracts'' and ``implements rules enabling individuals to coordinate for decentralized governance of an organization.''

\paragraph{Lummis-Gillibrand Responsible Financial Innovation Act}\label{fedbill}

In June 2022, Senators Cynthia Lummis and Kirsten Gillibrand introduced S.4356,\footnote{\url{https://www.congress.gov/bill/117th-congress/senate-bill/4356}} which has a section titled ``Decentralized autonomous organizations.'' Within this section, the bill provides a legal definition for a DAO with three qualifiers:

\begin{enumerate}
	\item ``An organization which utilizes smart contracts (as defined in section 9801 of title 31, United States Code) to effectuate collective action for a business, commercial, charitable, or similar entity,
	\item Governance of which is achieved primarily on a distributed basis, and,
	\item Which is properly incorporated or organized under the laws of a State or foreign jurisdiction as a decentralized autonomous organization, cooperative, foundation, or any similar entity.''
\end{enumerate}

Another observation from the bill's definitions is that it does not include any references to the term `blockchain'. Instead, it utilizes the `distributed ledger technology' phrase that is also found in the Tennessee DAO Law.

Lastly, the bill also includes several sections relating to stablecoins. These will be discussed more in Section~\ref{stable}. But in regard to the bill's definition for a `payment stablecoin', these are defined as digital assets with six qualifiers:

\begin{enumerate}[(A)]
	\item ``Redeemable, on demand, on a 1-to-1 basis for instruments denominated in United States dollars;
	\item Defined as legal tender under section 5103 or under the laws of a foreign country (excluding digital assets);
	\item Issued by a business entity;
	\item Accompanied by a statement from the issuer that the asset is redeemable, as specified in subparagraph (A), from the issuer or another identified person;
	\item Backed by 1 or more financial assets (excluding other digital assets), consistent with subparagraph (A); and
	\item Intended to be used as a medium of exchange.''
\end{enumerate}

\subsubsection{KYC, AML, and Digital IDs}\label{kycaml}

Another major legal component for creating an MBS-DAO is compliance with Know-Your-Customer (KYC) and Anti-Money Laundering (AML) laws. In regard to any security tokens issued by an MBS-DAO, these would be subject to registration with the Securities and Exchange Commission.

The topics of KYC and AML relate to the concept of digital identity. Considering the necessity for local governments to be involved in the real estate tokenization process, this seems like a natural place for identity verification to begin. In regard to legal recognition of any real estate tokenization system, there would necessarily be collaboration between the property owner, local government, and the developers of the tokenization system. Perhaps one output of this collaboration could be a pseudonymous identifier that could be used to maintain privacy for property owners. Then perhaps that identifier could also serve as KYC and AML compliance in regard to the buying/selling of the MBS-DAO governance tokens.

There are many possible ways to envision the role of digital identities within the future of the financial system. On the Bitcoin blockchain, the pseudonymous address structure creates some degree of obfuscation between connecting digital wallets to real world identities. However, there certainly techniques for de-anonymizing types of wallets. See \citet{Yin2019} and \citet{Jawaheri2020} for some examples of this. 

With the Ethereum blockchain, the reusability of addresses makes an Ethereum address effectively a form of digital id. On their webpage,\footnote{\url{https://ethereum.org/en/decentralized-identity/}} it describes more about the nature of `decentralized identity' on the Ethereum network and lists several tools for creating and managing identity. For example, the top entry in the transaction list from Figure~\ref{citydaowashsale} of Appendix~\ref{citydaoappend} contains an example where both the buyer and seller had registered an ENS address to replace the hash-based address of their Ethereum account.\footnote{See \url{https://ens.domains/} for more info on Ethereum Name Service (ENS). In short, it is similar to registering a web domain; however, it allows for the site to have economic identity on the Ethereum blockchain.} 

Similarly, the Cardano blockchain has tools built around providing a human-readable name to an account or wallet. For example, ADA Handles are NFTs minted on the Cardano blockchain that specify a link to a particular handle.\footnote{\url{https://mint.handle.me/}} These can then be sold on secondary marketplaces just like other NFTs. Another project on Cardano focused on digital ids is Atala PRISM,\footnote{\url{https://atalaprism.io}} which is being developed by Input Output.\footnote{\url{https://iohk.io}} Their take on `self-sovereign identity' involves using the Cardano blockchain to issue digital ids (DIDs) along with tools for practical use cases of identity verification.

\subsubsection{Stablecoins}\label{stable}

Although many DeFi projects in crypto involve the use of volatile crypto assets, the use of stablecoins can potentially eliminate those risks by allowing transactions to be settled in a digital representation of U.S. Dollars. The general concept of a stablecoin is a digital asset that achieves a stable value relative to some other asset. Typically, the `other asset' is the U.S. Dollar, which effectively implies a digital dollar; however, it is possible to create stablecoins that mimic other assets, such as Euros, gold, or even other crypto-assets. The \citet{PWGFM2021} define these as:
\begin{quote}
	``Stablecoins are digital assets that are designed to maintain a stable value relative to a national currency or other reference assets.''
\end{quote}

\citet{Fidelity2023} identifies four different types of stablecoins: (1) fiat-backed, (2) commodity-backed, (3) crypto-backed, and (4) algorithmic stablecoins. The first of these categories is the most straightforward where the value of the stablecoin is pegged to some fiat currency, and the backing of that peg is achieved by maintaining reserves in that currency. This is the category that most of the largest USD stablecoins fall into, such as Tether (USDT)\footnote{\url{https://tether.to/en/}} and USDC,\footnote{\url{https://www.circle.com/en/usdc}} which have combined to issue more than \$100 billion as of late 2023. With these fiat-backed stablecoins, the general idea is for each token (digital dollar) to be backed by an actual dollar (or very low-risk asset, such as short-term Treasury bonds) housed in a regulated financial institution.

The second type, commodity-backed stablecoins, is conceptually similar to fiat-backed stablecoins; however, the pegged asset is a commodity, rather than a fiat currency. The most established examples of this type are gold-backed stablecoins, such as Tether Gold (XAUT)\footnote{\url{https://gold.tether.to/}} and PAX Gold (PAXG),\footnote{\url{https://paxos.com/paxgold/}} which have issued over \$900 million in gold-backed stablecoins as of late 2023.

In the category of crypto-backed stablecoins, these are often referred to as ``wrapped'' crypto-assets and effectively enable cross-blockchain representations of crypto-assets. The largest among these is Wrapped Bitcoin (WBTC), which is an ERC-20 token that transacts on the Ethereum blockchain.\footnote{\url{https://wbtc.network/}} Each WBTC is backed by real BTC on the Bitcoin blockchain, which can be verified on the public ledger. This effectively allows Ethereum applications to incorporate BTC into DeFi platforms. However, given the reliance on smart contracts for such cross-chain `bridging', this introduces additional risks of potential contract vulnerabilities.\footnote{See \url{https://ethereum.org/en/bridges} for more on blockchain bridges, use cases, and associated risks.}

The last category, algorithmic stablecoins, is the most controversial, particularly following the collapse of the TerraUSD stablecoin in May 2022.\footnote{See \citet{Briola2023} and \citet{Lee2023} for more details about this particular case.} The general concept of this type of stablecoin is to create an overcollateralized system where deviations from the peg will automatically trigger actions to remedy the deviation. As a simple example, if the value of the stablecoin increases above the peg, the protocol will issue more of the token to increase supply and bring the value back down. If the price drops below the peg, then a mechanism to `burn' the tokens is triggered to reduce the supply and drive the price back up. For a more formal exploration of algorithmic stablecoins, see \citet{Zahnentferner2023}, which proposed a stablecoin called Djed that was launched on the Cardano blockchain in 2023.

\subsection{Smart Contract Programming Languages}\label{programming}

When discussing the various programming languages that could be used to develop the smart contracts necessary to construct an MBS-DAO, a natural starting point is to discuss Bitcoin and its scripting system that allows for the creation of wallets with programmatic conditions to spend the bitcoins. One of the simpler (and earliest) types of smart contracts is the creation of multi-signature (multisig) wallets, which require a pre-specified number of digital signatures to authorize spending out of the wallet. This type of multisig functionality forms the foundation for building more complex smart contracts, including DAOs and Layer-Two blockchain protocols.\footnote{See \citet{Gudgeon2020} for more details on Layer-Two solutions.}

On the Bitcoin blockchain, the basic smart contract functionality for multisig wallets was enabled with BIP 16 back in 2012.\footnote{Bitcoin Improvement Proposals (BIPs) are documents for proposing updates to the Bitcoin protocol. See \url{https://en.bitcoin.it/wiki/Bitcoin_Improvement_Proposals} for more info.} These P2SH (pay-to-script-hash) addresses begin with a `3', which distinguishes them from the original P2PKH (pay-to-pub-key-hash) addresses, which start with a `1'. Then, as part of the SegWit upgrade in 2017,\footnote{The Segregated Witness (SegWit) protocol upgrade for Bitcoin involved the adoption of several BIPs. For the Bech32 address format, the revelant BIPs are BIP 0173 and BIP 0350.} the Bech32 address format (beginning with `bc1') was introduced, which also allows for scripting.

On the Ethereum blockchain, the Solidity programming language has become the primary smart contracting language.\footnote{See \url{https://soliditylang.org/} and \url{https://github.com/ethereum/solidity} for more info.} These contracts are behind the vast majority of decentralized applications (dapps) running on Ethereum, such as CityDAO, Aave, Compound, and MakerDAO. The Ethereum website aggregates many tools and programs for learning Solidity.\footnote{See \url{https://ethereum.org/en/developers/learning-tools/} for details.} As of late 2023, the Solidity documentation includes examples for `Voting', `Blind Auction', `Safe Remote Purchase', and `Micropayment Channel'.\footnote{\url{https://docs.soliditylang.org/en/v0.8.23/solidity-by-example.html}}

On the Cardano blockchain, smart contract functionality was first released in late 2021 with the Alonso hardfork. During the years that followed, additional tooling and the development of smart contract standards have been released for the blockchain. The Haskell programming language\footnote{\url{https://www.haskell.org/}} forms the foundation of the Cardano blockchain, and the Plutus programming language,\footnote{\url{https://github.com/input-output-hk/plutus}} is built on that foundation to enable the development of smart contracts. The next layer is the development of financial smart contracts, which is the focus of the Marlowe domain-specific language.\footnote{\url{https://marlowe.iohk.io/}} In other words, Marlowe is a set of pre-designed Plutus contract templates that are tailored for financial contracts. 

As of late 2023, the Marlowe Contract Gallery\footnote{\url{https://docs.marlowe.iohk.io/docs/getting-started/contract-gallery}} includes examples for various NFT contracts, including `Sale of token for ada', `Sale of token for stablecoins', `Sale of token with royalties', `Shared ownership of an NFT', `NFT used as collateral for a loan', and `Auction of an NFT'. Additionally, the gallery contains several other relevant examples, such as `ACTUS contract for principal at maturity', `Coupon bond with guarantor', and `A geolocated Marlowe contract'.

\subsubsection{Smart Contract Oracles}\label{oracles}

The concept of a smart contract oracle is a system through which \emph{on-chain} smart contracts can incorporate \emph{off-chain} data/information. In regard to the scheduled cash flows of a fixed rate mortgage, the amount of off-chain information necessary to construct the mortgage smart contract is minimal. However, if we consider an adjustable rate mortgage, the dynamic interest rate and cash flows are a function of some baseline rate plus an additional spread. Although the spread calculation would likely be specified at origination, oracles could be key to transmitting the data regarding the dynamic baseline rate into the smart contract.

One popular project focused on the development of smart contract oracle services is Chainlink.\footnote{\url{https://chain.link/}} A search for `real estate' on the Chainlink Blog produces several posts focused on the topics of tokenized real estate and supplementing on-chain conditions with external data feeds from off-chain sources. For example, \citet{Chainlink2023} explores various aspects of real estate tokenization and demonstrates how this concept relates to each of the three steps that were discussed in Section~\ref{implementation}.

\subsection{Conclusion}\label{conclusion}

A significant contributor to the disruptive decline in the value of MBSs during the mid-2000s was the inability of investors to obtain timely information regarding cash flows. In this paper, we have explored the feasibility of creating a mortgage-backed security decentralized autonomous organization (MBS-DAO). How close are we to being able to instantaneously know the behavior of borrowers? Given that several U.S. states have already established DAOs as legal business structures (including proposed federal legislation), and given that ACTUS has already provided standards for amortizing securities, we are getting close. 

An MBS-DAO would primarily hold digital tokens that represent debt secured by real estate. This paper explored three key steps toward the creation of an MBS-DAO. (1) The first step is the tokenization of the underlying real estate. If a property's title can be represented with an NFT, then a mortgage smart contract could hold the NFT while the contract terms are being executed. (2) The next step toward replicating the traditional MBS structure is the tokenization of the cash flow rights to the mortgage contract. These tokens would effectively be mortgage security tokens. (3) Finally, an MBS-DAO could form a portfolio of those mortgage security tokens and issue MBS-DAO tokens. The MBS-DAO smart contract would redirect the cash flows generated by the portfolio to the tokenholders according to some predetermined algorithm. Those MBS-DAO tokens could also double as governance tokens that allow investors to participate directly in the governance of the DAO.

While there are still some components to this process yet to be realized, particularly around gaining legal recognition for tokenized representations of real world assets, this paper provides a review of the relevant streams of literature, an exploration into applying the ACTUS financial contract standards to design an MBS-DAO, and a discussion of several legal and technical considerations for such a pursuit. Throughout those sections, we discussed several examples of projects experimenting in the space of real estate tokenization: CityDAO, GoKey DAO, and Liquid Mortgage, with the last claiming to have reduced the reporting lags for MBS from 55 days to 30 minutes.

%%%%%%%%%%%%%%%%%%%%%%%%%%%%%%%%%%%%%%%%%%%%%%%%%%%%%%%%

\pagebreak
\setstretch{1.15}
\bibliography{mbsdaobib}

\begin{thebibliography}{}

\bibitem[\protect\citeauthoryear{{111th Congress of the U.S.}}{{111th Congress
  of the U.S.}}{2010}]{DoddFrank2010}
{111th Congress of the U.S.} (2010, July).
\newblock {H.R.4173 - Dodd-Frank Wall Street Reform and Consumer Protection
  Act}.
\newblock \url{https://www.congress.gov/bill/111th-congress/house-bill/4173}.

\bibitem[\protect\citeauthoryear{Ackerman and Davidson}{Ackerman and
  Davidson}{2019}]{Ackerman2019}
Ackerman, A. and K.~Davidson (2019, September).
\newblock {Fannie, Freddie to Retain Earnings}.
\newblock \emph{Wall Street Journal}.
\newblock
  \url{https://www.wsj.com/articles/fannie-freddie-to-retain-earnings-11569851912}.

\bibitem[\protect\citeauthoryear{Ackerman and Kendall}{Ackerman and
  Kendall}{2021}]{Ackerman2021}
Ackerman, A. and B.~Kendall (2021, June).
\newblock {Biden Administration Removes Fannie, Freddie Overseer After Court
  Ruling}.
\newblock \emph{Wall Street Journal}.
\newblock
  \url{https://www.wsj.com/articles/supreme-court-issues-mixed-ruling-on-government-seizure-of-fannie-freddie-profits-11624459222}.

\bibitem[\protect\citeauthoryear{Arshadi}{Arshadi}{2023}]{Arshadi2023}
Arshadi, N. (2023).
\newblock {Blockchain, Corporate Structure, and Financial Intermediation}.
\newblock {\em Technology \& Innovation\/}~{\em 23}, 1--22.

\bibitem[\protect\citeauthoryear{Briola, Vidal-Tom{\'a}s, Wang, and
  Aste}{Briola et~al.}{2023}]{Briola2023}
Briola, A., D.~Vidal-Tom{\'a}s, Y.~Wang, and T.~Aste (2023).
\newblock {Anatomy of a Stablecoin's failure: The Terra-Luna case}.
\newblock {\em Finance Research Letters\/}~{\em 51}, 103358.

\bibitem[\protect\citeauthoryear{Bundi}{Bundi}{2020}]{ACTUS}
Bundi, N. (2020).
\newblock {ACTUS: The algorithmic representation of financial contracts}.
\newblock \url{https://www.actusfrf.org/techspecs}.

\bibitem[\protect\citeauthoryear{Chainlink}{Chainlink}{2023}]{Chainlink2023}
Chainlink (2023, November).
\newblock {What Is Tokenized Real Estate?}
\newblock \emph{Chainlink Education}.
\newblock \url{https://chain.link/education-hub/tokenized-real-estate}.

\bibitem[\protect\citeauthoryear{Crow and Ostroff}{Crow and
  Ostroff}{2021}]{Crow2021}
Crow, K. and C.~Ostroff (2021, March).
\newblock {Beeple NFT Fetches Record-Breaking \$69 Million in Christie's Sale}.
\newblock \emph{Wall Street Journal}.
\newblock
  \url{https://www.wsj.com/articles/beeple-nft-fetches-record-breaking-69-million-in-christies-sale-11615477732}.

\bibitem[\protect\citeauthoryear{de~Jong, Kooijmans, and Koudijs}{de~Jong
  et~al.}{2023}]{Jong2023}
de~Jong, A., T.~Kooijmans, and P.~Koudijs (2023).
\newblock {Plantation Mortgage-Backed Securities: Evidence from Surinam in the
  Eighteenth Century}.
\newblock {\em The Journal of Economic History\/}~{\em 83\/}(3), 874--911.

\bibitem[\protect\citeauthoryear{Dombrowski}{Dombrowski}{2024}]{Dombrowski2024a}
Dombrowski, T. (2024).
\newblock {From Plantations to Blockchains: A Review and Synthesis of the MBS
  and DeFi Literatures}.
\newblock {\em Working Paper\/}.

\bibitem[\protect\citeauthoryear{Ewendt}{Ewendt}{2018}]{Ewendt2018}
Ewendt, M. (2018).
\newblock {Leveraging Blockchain Technology in Property Records: Establishing
  Trust in a Risk-Filled Market}.
\newblock {\em North Carolina Journal of Law \& Technology\/}~{\em 19\/}(4),
  99--126.

\bibitem[\protect\citeauthoryear{Ferreira}{Ferreira}{2021}]{Ferreira2021}
Ferreira, I. (2021, July).
\newblock {Decentralized systems and methods for managing loans and securities.
  (U.S. Patent No.: 11,068,978)}.
\newblock U.S. Patent and Trademark Office.
  \url{https://image-ppubs.uspto.gov/dirsearch-public/print/downloadPdf/11068978}.

\bibitem[\protect\citeauthoryear{Fidelity}{Fidelity}{2023}]{Fidelity2023}
Fidelity (2023, August).
\newblock {What is a stablecoin?}
\newblock
  \url{https://www.fidelity.com/learning-center/trading-investing/what-is-a-stablecoin}.

\bibitem[\protect\citeauthoryear{Freyermuth, Odinet, and Tosato}{Freyermuth
  et~al.}{2023}]{Freyermuth2023}
Freyermuth, R.~W., C.~K. Odinet, and A.~Tosato (2023).
\newblock {Crypto in Real Estate Finance}.
\newblock {\em Alabama Law Review\/}~{\em 75\/}(1), 93--156.

\bibitem[\protect\citeauthoryear{Gaffney}{Gaffney}{2023}]{Gaffney2023}
Gaffney, P. (2023, November).
\newblock {Tokenization and Real-World Assets Take Center Stage}.
\newblock \emph{CoinDesk}.
\newblock
  \url{https://www.coindesk.com/business/2023/11/22/tokenization-and-real-world-assets-take-center-stage/}.

\bibitem[\protect\citeauthoryear{Ghent and Kudlyak}{Ghent and
  Kudlyak}{2011}]{Ghent2011}
Ghent, A.~C. and M.~Kudlyak (2011).
\newblock {Recourse and Residential Mortgage Default: Evidence from US States}.
\newblock {\em The Review of Financial Studies\/}~{\em 24\/}(9), 3139--3186.

\bibitem[\protect\citeauthoryear{Goldberg and Sch{\"a}r}{Goldberg and
  Sch{\"a}r}{2023}]{Goldberg2023}
Goldberg, M. and F.~Sch{\"a}r (2023).
\newblock {Metaverse governance: An empirical analysis of voting within
  Decentralized Autonomous Organizations}.
\newblock {\em Journal of Business Research\/}~{\em 160}, 113764.

\bibitem[\protect\citeauthoryear{Gudgeon, Moreno-Sanchez, Roos, McCorry, and
  Gervais}{Gudgeon et~al.}{2020}]{Gudgeon2020}
Gudgeon, L., P.~Moreno-Sanchez, S.~Roos, P.~McCorry, and A.~Gervais (2020).
\newblock {SoK: Layer-Two Blockchain Protocols}.
\newblock In J.~Bonneau and N.~Heninger (Eds.), {\em Financial Cryptography and
  Data Security}, Cham, pp.\  201--226. Springer International Publishing.

\bibitem[\protect\citeauthoryear{Higgins}{Higgins}{2017}]{Higgins2017}
Higgins, S. (2017, August).
\newblock {\$26 Million: Blockchain VR Project Decentraland Raises New Funding
  in ICO}.
\newblock CoinDesk.
\newblock
  \url{https://www.coindesk.com/markets/2017/08/18/26-million-blockchain-vr-project-decentraland-raises-new-funding-in-ico/}.

\bibitem[\protect\citeauthoryear{Jawaheri, Sabah, Boshmaf, and Erbad}{Jawaheri
  et~al.}{2020}]{Jawaheri2020}
Jawaheri, H.~A., M.~A. Sabah, Y.~Boshmaf, and A.~Erbad (2020).
\newblock {Deanonymizing Tor hidden service users through Bitcoin transactions
  analysis}.
\newblock {\em Computers \& Security\/}~{\em 89}, 101684.

\bibitem[\protect\citeauthoryear{Koronczok}{Koronczok}{2019}]{Koronczok2019}
Koronczok, M. (2019).
\newblock {The New ``Chain'' of Title: How Blockchain Will Affect Land Title
  Research, Recordation, and Insurance}.
\newblock {\em Texas A\&M Journal of Property Law\/}~{\em 5\/}(3), 401--420.

\bibitem[\protect\citeauthoryear{Lee, Lee, and Lee}{Lee et~al.}{2023}]{Lee2023}
Lee, S., J.~Lee, and Y.~Lee (2023).
\newblock {Dissecting the Terra-LUNA crash: Evidence from the spillover effect
  and information flow}.
\newblock {\em Finance Research Letters\/}~{\em 53}, 103590.

\bibitem[\protect\citeauthoryear{Nagl, Nagl, R\"{o}sch, Sch\"{a}fers, and
  Freybote}{Nagl et~al.}{2023}]{Nagl2023}
Nagl, C., M.~Nagl, D.~R\"{o}sch, W.~Sch\"{a}fers, and J.~Freybote (2023).
\newblock {Time Varying Dependences Between Real Estate Crypto, Real Estate and
  Crypto Returns}.
\newblock {\em Journal of Real Estate Research\/}.

\bibitem[\protect\citeauthoryear{Nakamoto}{Nakamoto}{2008}]{Nakamoto2008}
Nakamoto, S. (2008).
\newblock {Bitcoin: A Peer-to-Peer Electronic Cash System}.
\newblock \url{https://bitcoin.org/bitcoin.pdf}.

\bibitem[\protect\citeauthoryear{{Office of Financial Research}}{{Office of
  Financial Research}}{2022}]{OFR2022}
{Office of Financial Research} (2022, November).
\newblock {OFR Expands Its Financial Instrument Reference Database to Help
  Identify Inconsistencies in Financial Terms}.
\newblock
  \url{https://www.financialresearch.gov/press-releases/2022/11/07/ofr-enhances-fird/}.

\bibitem[\protect\citeauthoryear{{President's Working Group on Financial
  Markets}, {the Federal Deposit Insurance Corporation}, and {the Office of the
  Comptroller of the Currency}}{{President's Working Group on Financial
  Markets} et~al.}{2021}]{PWGFM2021}
{President's Working Group on Financial Markets}, {the Federal Deposit
  Insurance Corporation}, and {the Office of the Comptroller of the Currency}
  (2021, November).
\newblock {Report on Stablecoins}.
\newblock
  \url{https://home.treasury.gov/system/files/136/StableCoinReport_Nov1_508.pdf}.

\bibitem[\protect\citeauthoryear{Saari, Vimpari, and Junnila}{Saari
  et~al.}{2022}]{Saari2022}
Saari, A., J.~Vimpari, and S.~Junnila (2022).
\newblock Blockchain in real estate: Recent developments and empirical
  applications.
\newblock {\em Land Use Policy\/}~{\em 121}, 106334.

\bibitem[\protect\citeauthoryear{Spielman}{Spielman}{2016}]{Spielman2016}
Spielman, A. (2016).
\newblock {Blockchain: Digitally Rebuilding the Real Estate Industry}.
\newblock mathesis, Massachusetts Institute of Technology.
\newblock \url{https://dspace.mit.edu/handle/1721.1/106753}.

\bibitem[\protect\citeauthoryear{Vasile}{Vasile}{2021}]{Vasile2021}
Vasile, I. (2021, November).
\newblock {Everything You Need To Know About Initial Stake Pool Offerings
  (ISPO)}.
\newblock \emph{BeInCrypto}.
\newblock \url{https://beincrypto.com/learn/initial-stake-pool-offerings/}.

\bibitem[\protect\citeauthoryear{Yermack}{Yermack}{2017}]{Yermack2017}
Yermack, D. (2017).
\newblock {Corporate Governance and Blockchains}.
\newblock {\em Review of Finance\/}~{\em 21\/}(1), 7--31.

\bibitem[\protect\citeauthoryear{Yin, Langenheldt, Harlev, Mukkamala, and
  Vatrapu}{Yin et~al.}{2019}]{Yin2019}
Yin, H. H.~S., K.~Langenheldt, M.~Harlev, R.~R. Mukkamala, and R.~Vatrapu
  (2019).
\newblock {Regulating Cryptocurrencies: A Supervised Machine Learning Approach
  to De-Anonymizing the Bitcoin Blockchain}.
\newblock {\em Journal of Management Information Systems\/}~{\em 36\/}(1),
  37--73.

\bibitem[\protect\citeauthoryear{yob Nam and Oh}{yob Nam and
  Oh}{2021}]{Nam2021}
yob Nam, T. and S.~Oh (2021).
\newblock Non-recourse mortgage law and housing speculation.
\newblock {\em Journal of Banking \& Finance\/}~{\em 133}, 106292.

\bibitem[\protect\citeauthoryear{Zahnentferner, Kaidalov, Etienne, and
  D{\'i}az}{Zahnentferner et~al.}{2023}]{Zahnentferner2023}
Zahnentferner, J., D.~Kaidalov, J.-F. Etienne, and J.~D{\'i}az (2023).
\newblock {Djed: A Formally Verified Crypto-Backed Autonomous Stablecoin
  Protocol}.
\newblock In {\em 2023 IEEE International Conference on Blockchain and
  Cryptocurrency (ICBC)}.

\bibitem[\protect\citeauthoryear{Zhao, Ai, Lai, Luo, and Benitez}{Zhao
  et~al.}{2022}]{Zhao2022}
Zhao, X., P.~Ai, F.~Lai, X.~R. Luo, and J.~Benitez (2022).
\newblock Task management in decentralized autonomous organization.
\newblock {\em Journal of Operations Management\/}~{\em 68\/}(6--7), 649--674.

\end{thebibliography}
\bibliographystyle{chicago}

\pagebreak

\appendix

\begin{landscape}
	\section{ACTUS Financial Contract Taxonomy}\label{actustaxappend}
	\begin{figure}[htbp]
		\centering
		\vspace{-2em}
		\includegraphics[height=0.75\textwidth]{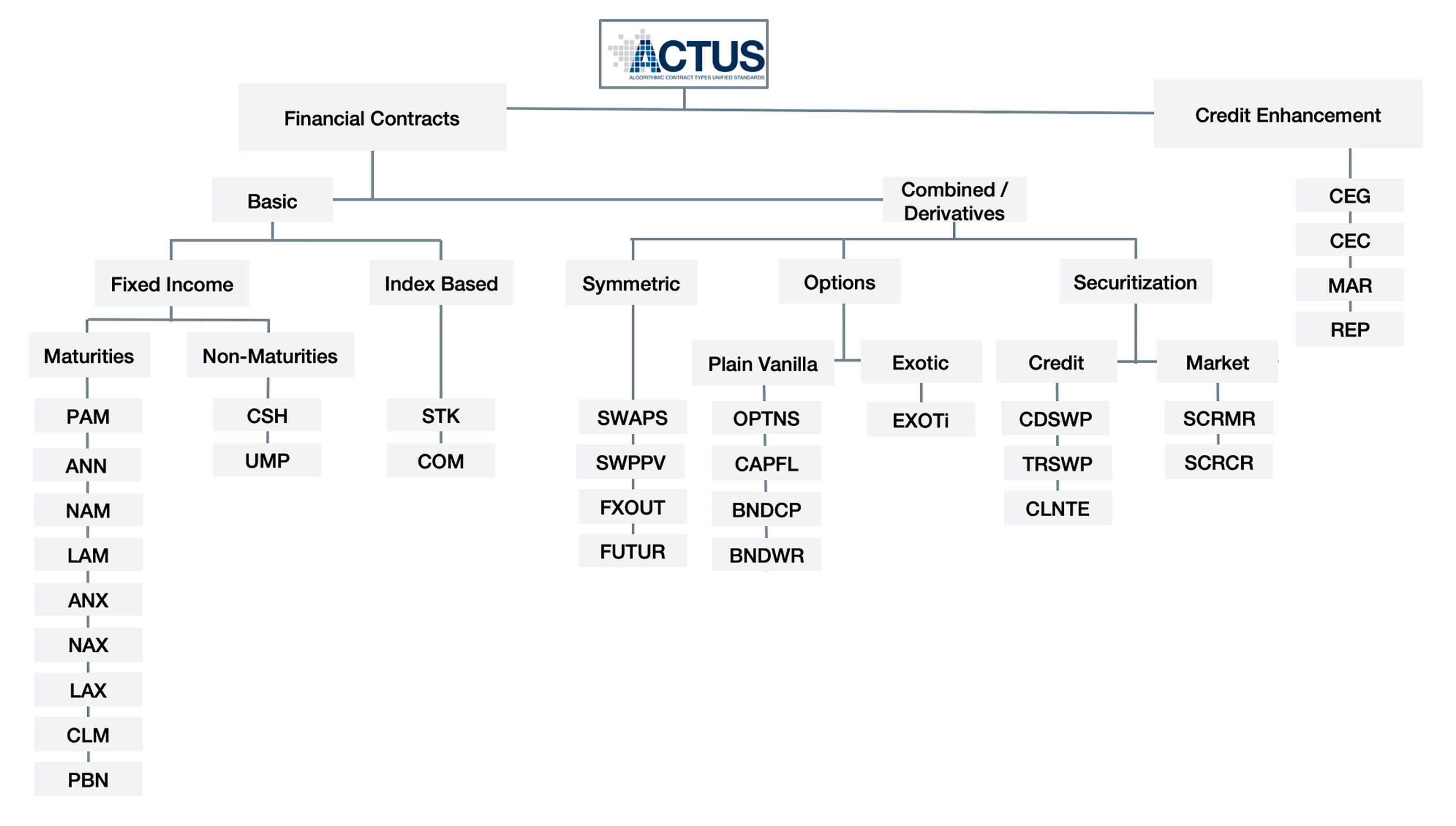}
		\caption{Algorithmic Contract Types Unifed Standards (ACTUS) Taxonomy -- \url{https://www.actusfrf.org/taxonomy}}
		\label{actustaxonomy}
	\end{figure}
\end{landscape}

\pagebreak

\section{CityDAO NFT Trading Activity}\label{citydaoappend}

Below are several screenshots from OpenSea\footnote{\url{https://opensea.io}} displaying trends for each of the first two NFT issues from CityDAO: Citizenship NFTs and Parcel-0 NFTs. After a brief review of the price and volume trends for each NFT project, an example is provided for idenifying wash sales from the blockchain transaction data, which explains the outliers in Figure~\ref{citydaoparcel0fig}.

\subsection{CityDAO Citizenship NFTs}

\begin{figure}[htbp]
	\centering
	\includegraphics[width=\textwidth]{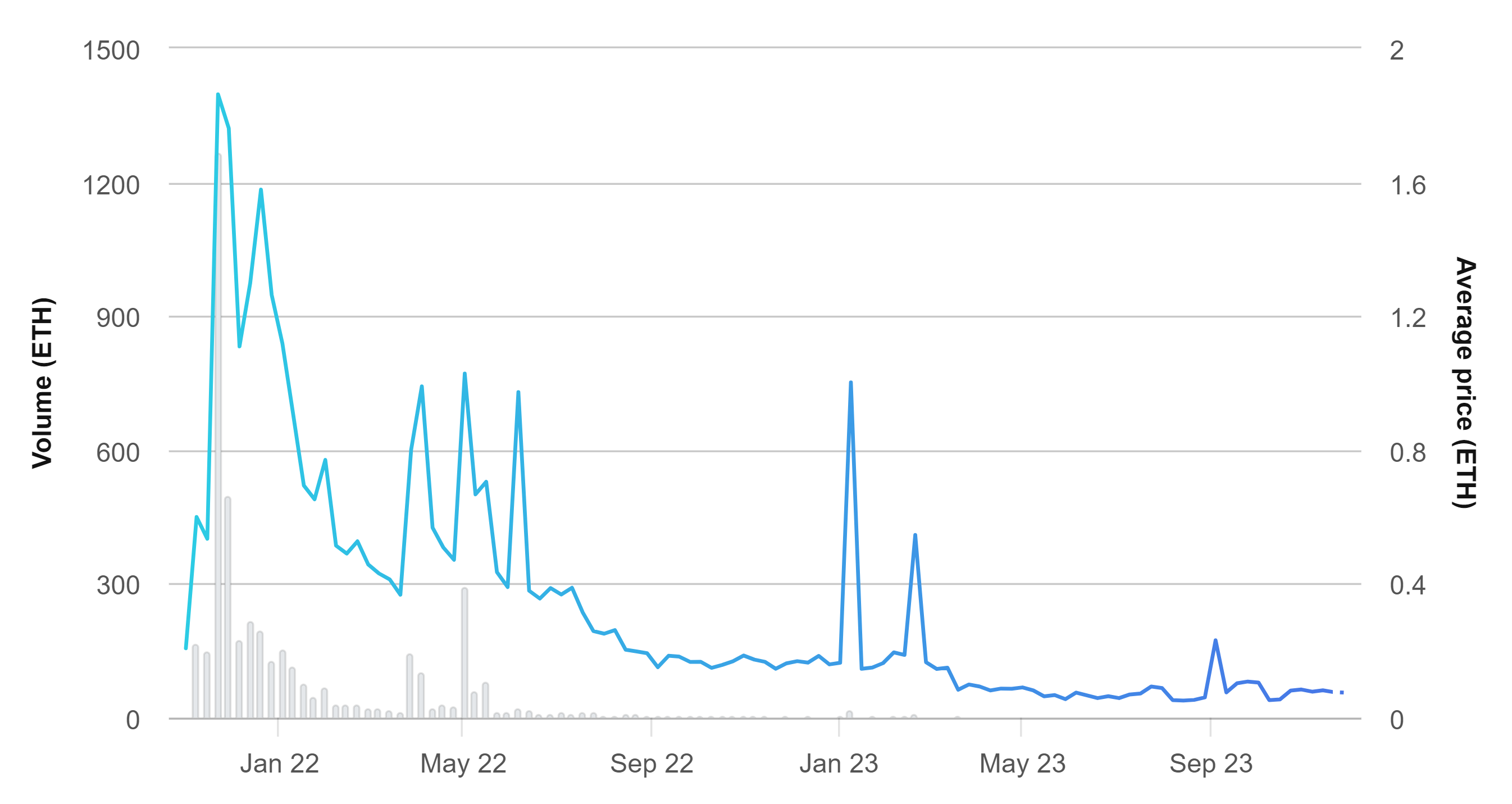}
	\caption{CityDAO Citizenship NFT price and volume chart from OpenSea \\ \url{https://opensea.io/collection/cdao/activity}}
	\label{citydaocitizenfig}
\end{figure}

First, in Figure~\ref{citydaocitizenfig}, presents the price and volume chart for the CityDAO Citizenship NFTs since their issuance in late 2021. This price (line) component of the chart tracks the average sale price using the on-chain transactions data. Similar to the general trends of the broad NFT markets, the declining value of the tokens is similar to most other NFT projects over this period.

Another observation from the price series for the Citizenship NFTs in Figure~\ref{citydaocitizenfig} is the occasional spikes that deviate from the general downward trend. These can be explained by the various tiers of the NFTs in this offering. The 10,051 NFTs in this project are split across three tiers: 10,000 Citizen NFTs,\footnote{\url{https://city.mirror.xyz/TRMrm5r5CYtYwaG3G4jG125pBv6dyuiTZ2FkBqskTBo}} 50 Founding Citizen NFTs,\footnote{\url{https://city.mirror.xyz/zoK5OaIdxYPqTmiK4WcGtR4VKXKDYnBjTjjo892YMmM}} and a unique 1-of-1 First Citizen NFT.\footnote{\url{https://city.mirror.xyz/CmZyzy3tuOu4TPUnfsYaR1Zoub1yiej6ZoOEuoayDQM}} Those rarer tiers come with more explicit governance rights, and the increased scarcity typically demands a higher price. Thus, the price spikes throughout 2023 can be attributed to the sale of a Founding Citizen token.\footnote{To verify this, \url{https://opensea.io/collection/cdao/activity} also lists all on-chain transactions.}

\subsection{CityDAO Parcel Zero NFTs}

\begin{figure}[htbp]
\centering
\includegraphics[width=\textwidth]{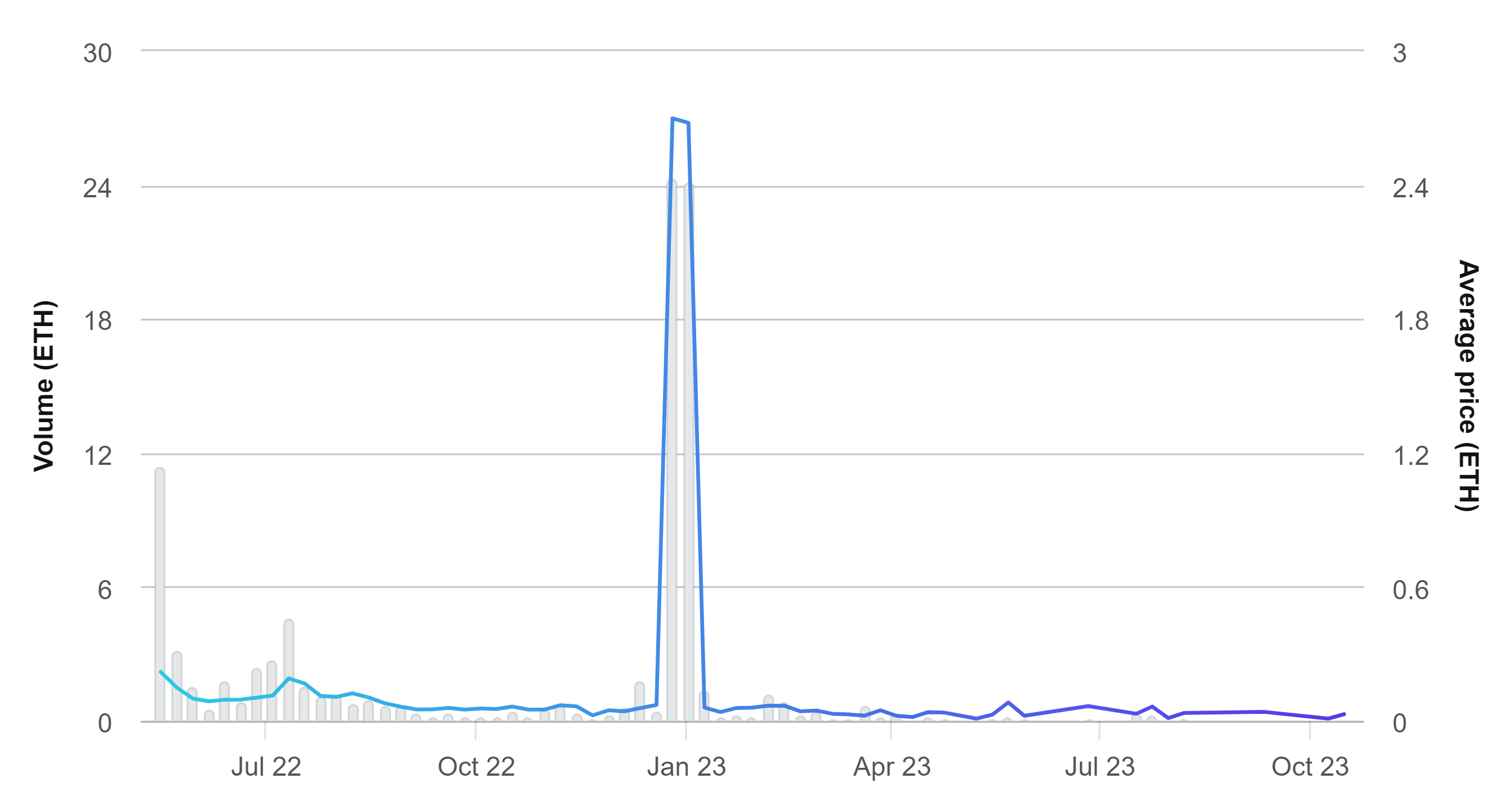}
\caption{CityDAO Parcel-0 NFT price and volume chart from OpenSea \\ \url{https://opensea.io/collection/citydao-parcel-0/activity}}
\label{citydaoparcel0fig}
\end{figure}

The other NFT issuance from CityDAO is that of the Parcel Zero NFTs. These each contain metadata that describes a particular plot from the real-world Parcel Zero property located in Wyoming. Of note in Figure~\ref{citydaoparcel0fig} is the large spike/outlier at the end of 2022 and beginning of 2023.

\begin{figure}[htbp]
	\centering
	\includegraphics[width=\textwidth]{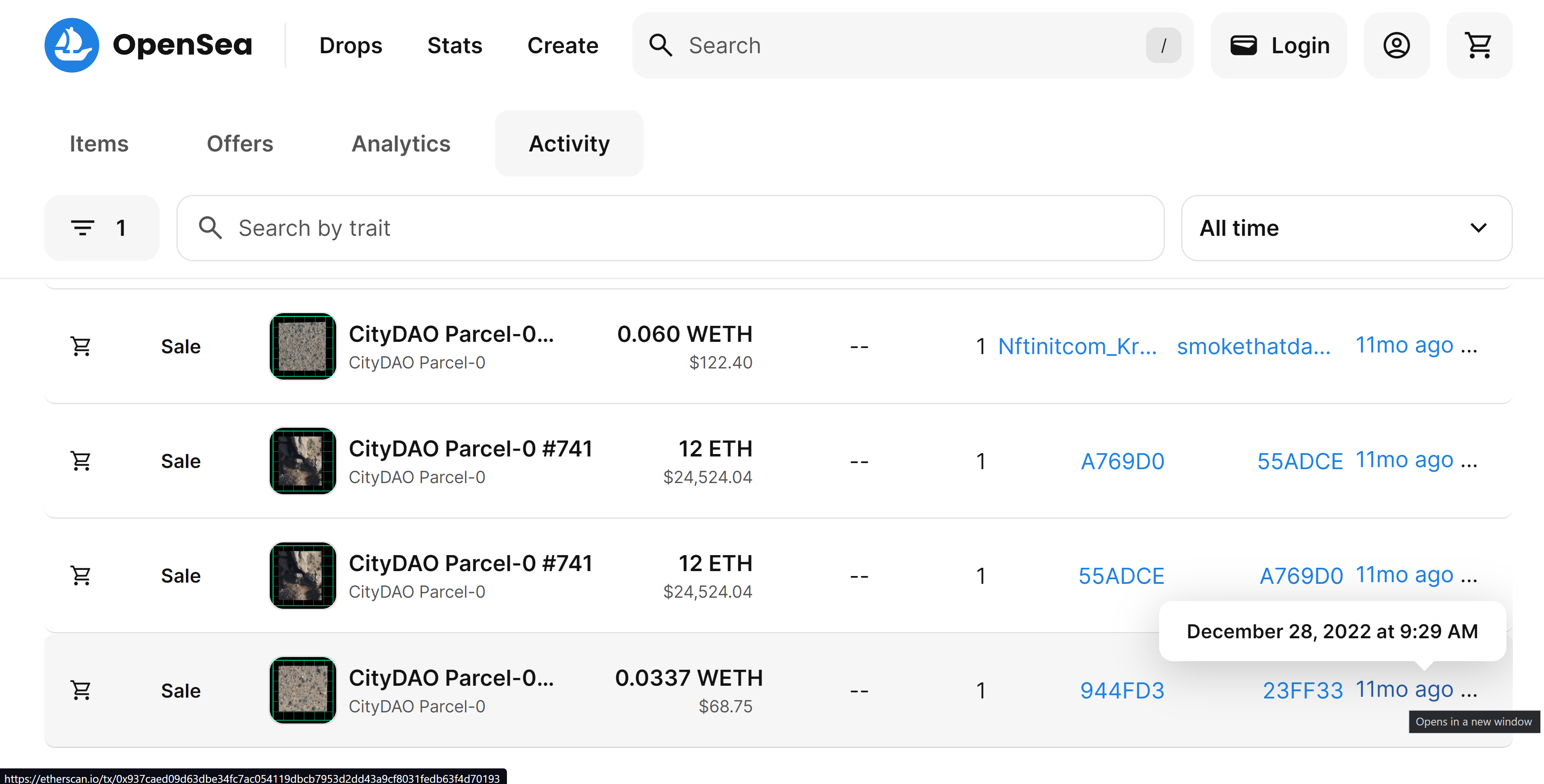}
	\caption{Example of a pair of wash sale transactions for a CityDAO Parcel Zero NFT \\ \url{https://opensea.io/collection/citydao-parcel-0/activity}}
	\label{citydaowashsale}
\end{figure}

A deeper investigation into the transaction data allows us to verify that these outliers are from wash sales where an over-priced purchase was reversed within a couple minutes. Figure~\ref{citydaowashsale} shows an example of how these transactions appear on OpenSea. Of note are the `From' and `To' columns, which are the two on the right side before the date column. From the pseudonymous Ethereum account identifiers, we can see that `55ADCE' sold the NFT to `A769D0' for 12 ETH in late 2022. Then three minutes later,\footnote{The three dots to the right of each entry link to the Etherscan page with more information from the blockchain, such as the block number, transaction fees, and details about the contract that facilitated the NFT sale. See \url{https://github.com/ProjectOpenSea/seaport} for more info about the Seaport protocol.} a reversal transaction back to `55ADCE' is executed.

This example of a wash sale may have been as simple as a couple test transfers by a member of CityDAO between two wallets. However, it expresses an important point in regard to relying on blockchain-based transaction data. Since anyone can create multiple pseudonymous Ethereum accounts quite easily, the identification of wash trading can be far more complex in practice. Thus, measures of trading volume can potentially include non-trivial amounts of wash sale activity.

\pagebreak

\section{Wyoming DAO-LLC Organization Form Instructions}\label{wyodaoappend}
\begin{figure}[htbp]
	\vspace{-1em}
	\centering
	\caption{Retrieved from \url{https://sos.wyo.gov/Forms/default.aspx}}
	

\section*{Supplementary material}

# C  Wyoming DAO-LLC Organization Form Instructions

Figure 7: Retrieved from `https://sos.wyo.gov/Forms/default.aspx`

## Decentralized Autonomous Organization Limited Liability Company Instructions

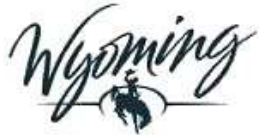

Wyoming Secretary of State
Herschler Building East, Suite 101 ◆ 122 W 25th Street ◆ Cheyenne, WY 82002-0020
307.777.7311 ◆ Business@wyo.gov

**Before Filing Please Note:**

☐ Pursuant to W.S. 17-29-108 and W.S. 17-31-104(d), the name must include "DAO LLC" or a combination of the entity indicators for both DAO *and* LLC as follows:
   - The name must include, specifically, "DAO", or "LAO."
   - The name must include the words "Limited Liability Company," or its abbreviations "LLC," "L.L.C.," "Limited Company," "LC," "L.C.," "Ltd. Liability Company," "Ltd. Liability Co.," or "Limited Liability Co."

☐ Under the circumstance specified in W.S. 17-28-104(e), **an email address is required.**

☐ *Filing fee of $100.00.* Visa or MasterCard payment available for online filings only. To file online, visit: https://wyobiz.wyo.gov. Make check or money order payable to Wyoming Secretary of State for paper filings.

☐ **Annual reports are due every year** on the first day of the anniversary month of formation. If not paid within 60 days of the due date the entity will be subject to dissolution.

☐ **Processing time is up to 15 business days** following the date of receipt in our office.

☐ Please review the form prior to submission. **The Secretary of State's Office is unable to process incomplete forms.**

**Additional Contact Information:**

◆ **Department of Revenue** (Sales and Use Tax Information)
   - Ph. 307.777.5200 OR https://revenue.state.wy.us/
◆ **Department of Workforce Services** (Workers' Compensation or Unemployment Insurance)
   - Ph. 307.777.8650 OR http://www.wyomingworkforce.org/
◆ **Internal Revenue Service** (Tax ID Information)
   - https://www.irs.gov/Filing


	\label{wyodaoform}
\end{figure}

\end{document}